\documentclass[preprint,3p]{elsarticle}

\usepackage{natbib}
\usepackage{paralist}
\usepackage{rotating}
\usepackage{graphicx}
\usepackage[pdftex, colorlinks,
            linkcolor=black,
            citecolor=black,
            filecolor=black,
            urlcolor=black]{hyperref}

\usepackage{amssymb}

\journal{Information and Software Technology}

\begin{document}

\begin{frontmatter}



\title{Naming the Pain in Requirements Engineering: \\ A Design for a Global Family of Surveys and First Results from Germany }

\author[label1]{Daniel M\'{e}ndez Fern\'{a}ndez\corref{cor1}}
\ead{mendezfe@in.tum.de}
\author[label2]{Stefan Wagner}

\address[label1]{Software \& Systems Engineering, Institut f\"ur Informatik, Technische Universit\"at M\"unchen, Boltzmannstr.~3, 85748 Garching, Germany}

\address[label2]{Software Engineering Group, Institute of Software Technology, University of Stuttgart, Universit\"atsstr.~38, 70569 Stuttgart, Germany}

\cortext[cor1]{Corresponding author}

\begin{abstract}
\textbf{Context:} For many years, we have observed industry struggling in defining a high quality requirements engineering (RE) and researchers trying to understand industrial expectations and problems. Although we are investigating the discipline with a plethora of empirical studies, they still do not allow for empirical generalisations. 

\textbf{Objective:} To lay an empirical and externally valid foundation about the state of the practice in RE, we aim at a series of open and reproducible surveys that allow us to steer future research in a problem-driven manner.  

\textbf{Method:} We designed a globally distributed family of surveys in joint collaborations with different researchers and completed the first run in Germany. The instrument is based on a theory in the form of a set of hypotheses inferred from our experiences and available studies. We test each hypothesis in our theory and identify further candidates to extend the theory by correlation and Grounded Theory analysis.

\textbf{Results:} In this article, we report on the design of the family of surveys, its underlying theory, and the full results obtained from Germany with participants from 58 companies. The results reveal, for example, a tendency to improve RE via internally defined qualitative methods rather than relying on normative approaches like CMMI. We also discovered various RE problems that are statistically significant in practice. For instance, we could corroborate communication flaws or moving targets as problems in practice. Our results are not yet fully representative but already give first insights into current practices and problems in RE, and they allow us to draw lessons learnt for future replications.

\textbf{Conclusion:} Our results obtained from this first run in Germany make us confident that the survey design and instrument are well-suited to be replicated and, thereby, to create a generalisable empirical basis of RE in practice.
\end{abstract}


\end{frontmatter}


\section{Introduction}
\label{sec:Introduction}

Requirements engineering (RE) is a discipline that constitutes a holistic key to successful development projects as the elicitation, specification and validation of precise and stakeholder-appropriate requirements are critical determinants of software \& system quality~\citep{broy06_mbRE}. At the same time, RE is characterised by the involvement of interdisciplinary stakeholders and uncertainty as many aspects are not clear from the beginning of a project. Hence, RE is highly volatile and inherently complex by nature. 

Although the importance of a high quality RE and its continuos improvement have been recognised for many years, we can still observe industry struggling in defining and applying a high quality RE~\citep{MWLBC10}. The diversity of how RE is performed in various industrial environments, each having its particularities in the domains of application or the software process models used, dooms the discipline to be not only a process area difficult to improve, but also difficult to investigate for common practices and shortcomings. 

From a researcher's perspective, empirical research in RE thereby becomes a crucial and challenging task. It is crucial as empirical studies of any kind in RE, ranging from classical action research through observational studies to broad exploratory surveys, are necessary to understand the practical needs and improvement goals in RE, to steer problem-driven research and to empirically validate new research proposals~\citep{CDW12}. It is challenging, because we still need a solid empirical basis that allows for generalisations taking into account the human factors that influence the anyway hardly standardisable discipline like no other in software engineering. In consequence, qualitative research methods are gaining much attention~\citep{Seaman99}, and survey research has become an indispensable means to investigate RE especially in a broader context.

\subsection{Problem Statement}
Although we are confident about the value of survey research to understand practical needs and to distill improvement goals in RE, we still lack a solid empirical survey basis. The reason seems to lie in an ironically paradoxical circumstance: The appropriate design of a survey in RE and the statistically sound interpretation of the results going beyond purely observational, qualitative analyses and reasoning is very challenging, because we still lack empirically grounded theories in RE~\citep{CDW12}. In turn, we still lack such theories in RE as we still have no empirically sound survey basis. Available surveys in RE either investigate isolated techniques in application (small scope), or they focus on a small data population (single countries or companies) so that the findings of the surveys are hardly generalisable -- they cannot be viewed as representative. 

Yet missing is a series of grounded empirical investigations of practical problems and needs in RE that allows for empirical generalisations to steer future research in a problem-driven manner.

\subsection{Research Objective}

Our long-term research objective is to establish an open and externally valid set of empirical findings about practical problems and needs in RE that allows us to steer future research in a problem-driven manner. To this end, we want to conduct a continuously and independently replicated, globally distributed survey on RE that investigates the state of the practice and trends including industrial expectations, status quo, experienced problems and what effects those problems have. The work presented in this article forms the first steps in this direction.

\subsection{Contribution}
\label{sec:contribution}

In this article, we contribute the design of a globally distributed family of surveys on RE and the results obtained from its initial start in Germany. Our instrument relies on an initial theory obtained from available RE studies and used to generate hypotheses which we test during the results analysis. The results gathered from open questions are further used to already extend our initial theory using Grounded Theory analysis. Furthermore, we investigate patterns in statistically significant correlations to find further candidate hypotheses for the theory.

In~\cite{MW2013a}, we already published the design of the family of surveys and the used instrument, and we discussed preliminary results from its initial start in Germany. In this article, we extend the previously published material by contributing:

\begin{compactenum}
\item A detailed theory on which the improvement of requirements engineering practices can be based on. We infer our initial theory from our experiences and available empirical studies in RE. This theory includes for each research question a set of testable hypotheses about industrial expectations on good RE, the status quo in RE and its improvement, and contemporary problems in RE. We use the hypotheses in our theory to test our results during our analysis procedure.
\item The full results from Germany including the responses from 58 companies (out of 73 responses) and a detailed analysis of those results via:
\begin{compactitem}
\item A series of hypotheses tests w.r.t. to our theory. This allows a first statistically sound interpretation of the answers and an initial validation of the theory. 
\item A correlation analysis among all variables in the complete data set to investigate whether statistically significant patterns can be distilled.
\item An investigation of the implications specific problems in RE have. As this topic is barely investigated yet, we have no predefined expectations and, thus, we apply Grounded Theory (to potentially extend our initial theory).
\end{compactitem}
\item A set of lessons learnt from our initial run in Germany used to modify the process and the instrument for the replication that is already ongoing at the time of writing this article.
\end{compactenum} 

Our contribution is thus intended to serve two audiences: researchers that will participate in the survey replications and ones interested in the survey results.

\subsection{Terminology Used in the Survey}
\label{sec:terminology}
To ensure an unambiguous terminology for our paper, we explain the most frequently used terms to report on our survey. Table~\ref{tab:terminology} summarises for this purpose relevant terms and definitions.

\begin{table}[htb]
\caption{Terminology used in this paper.
\label{tab:terminology}}
\begin{center}
\begin{tabular}{rp{0.7\linewidth}}
\hline
\textbf{Artefact} & Deliverable that abstracts from contents of a specification document. It is used as input, output, or as an intermediate result of a process definition (see also~\cite{MPKB10}). Synonyms used: Work product, document.\\
\textbf{Grounded theory} &  Grounded Theory is a set of integrated concepts/hypotheses, which are systematically generated from qualitative and quantitative data (see~\cite{GS67, AHK11}).\\
\textbf{Replication study} & Reproduction of an empirical study with the purpose of empirical generalisation, i.e. continuous repetition of a study until reaching a saturation in the observations objectively leading to same or similar results as the replicated studies. Each replication is performed independently by different researchers (see~\cite{GJV10}). \\
\textbf{RE standard} & Standardised organisational requirements engineering (RE) model that abstracts from the idealised execution of the RE phase in a development project. It includes the description of the process (definition) to follow, the artefacts to be created, as well as roles involved~\cite{PS06}. Synonyms used: RE reference model, methodology~\cite{ISO24744} (taking a broader view)\\
\textbf{Theory} & ``A theory provides explanations and understanding in terms of basic concepts and underlying mechanisms, which constitute an important counterpart to knowledge of passing trends and their manifestation''~\cite{HDT07, WRH+12}. We use this term in the sense of a \emph{social theory} (see~\cite{Popper02, AHK11}) and refer to a set of falsifiable and testable statements/hypotheses having an inductive nature with no apodictic propositions.\\

\hline
\end{tabular}
\end{center}
\end{table}

\subsection{Outline}

The remainder of the article is as follows. In Sect.~\ref{sec:RelatedWork}, we introduce available contributions in the context of our research, which gaps are left open, and how we intent to close those gaps. In Sect.~\ref{sec:Theory}, we present our starting theory in the form of a series hypotheses for our areas of interested. In Sect.~\ref{sec:Design}, we present the design of the family of surveys, including the research questions based on the theory, the used methodology and instrumentation, and the data analysis and validity procedures. The results of the survey in Germany are reported in Sect.~\ref{sec:Results}. In Sect.~\ref{LessonsLearnt}, we summarise our lessons learnt and discuss those things we change to in future to facilitate the replications, before giving a summary of results and a concluding discussion in Sect.~\ref{sec:Conclusion}.

\section{Related Work}
\label{sec:RelatedWork}
As the topics of our survey touch many different areas in requirements engineering, it is impossible 
to comprehensively discuss all related work in that sense. Therefore, we focus on survey research
on RE and our own previously published material. 

\subsection{Survey Research on Requirements Engineering}

We classify RE survey research into two major areas: investigations of techniques and methods and 
investigations of general practices and contemporary phenomena in industry. Contributions that investigate 
techniques and methods analyse, for example, selected requirements phases and which techniques are suitable 
to support typical tasks in those phases. Zowghi et al.~\cite{ZC05}, for example, conducted a survey about which 
techniques support the elicitation phase. 
An exemplary survey on the choice of elicitation techniques was carried out by Carrizo et al.~\cite{CDJ08}. Studies like 
those reveal the effects of given techniques when applying them in practical contexts. 


Surveys on general practices and phenomena in industry include the well-known Chaos Report of the Standish Group, examining, inter alia, root causes for project failures of which most are to be seen in RE, such as missing user involvement. Whereas the report is known to have serious flaws in its design negatively affecting the validity of the results~\citep{EV10}, other studies, such as the (German) Success~study~\cite{success07}, conduct a similar investigation of German companies including a detailed and reproducible study design. Still, both surveys exclusively investigate failed projects and general causes at the level of overall software processes. 

The focus of those studies, however, does not support the investigation of contemporary phenomena and problems of RE 
in industry. Neill and Laplante~\cite{Neill:2003ui} conducted a survey on current practices in RE; for example, the used
requirements elicitation and modelling techniques. 
Nikula et al.~\cite{nikula2000sps} present a survey on RE at the organisational level of small and medium-size companies in Finland. Based on their findings, they inferred improvement goals, e.g., on optimising knowledge transfer. 
A study investigating the industrial reluctance on software process improvement was performed by Staples et al.~\cite{SNJABR07}. They discovered different reasons why organisations do not adopt normative improvement solutions as the ones dictated by CMMI and related frameworks (focussing on assessing and benchmarking companies rather than on conducting problem-driven, inductive improvements, see~\cite{NMJ09,PIGO08,MW2013a}). Exemplary reasons for a reluctance to normative improvement frameworks were the small company size where the respondents did not see clear benefit. 
A survey that directly focused on discovering problems in practical settings was performed by Hall et al.~\cite{HBR02}. They empirically underpin the problems discussed by Hsia et al.~\cite{HDK93} and investigated a set of critical organisational and project-specific problems, such as communication problems, inappropriate skills or vague requirements.

Solemon, Sahibuddin, and Abd Ghani~\cite{Solemon:2009tf} report on a survey on RE problems and practices in 
Malaysian software companies. They found several of the RE problems we also saw in our survey. 
Liu, Li, and Peng~\cite{Liu:jt} describe a survey conducted in China about practices and problems in RE. They
discuss several problems we also investigated but concentrate on China.
Verner et al.~\cite{Verner:2005vl} ran a survey in Australia and the USA. They concentrated on success factors
in RE and found good requirements, customer/user involvement, and effective requirements management to be
the best predictors of project success.


Although giving valuable insights into industrial environments, the discussed surveys do by now not allow for 
generalisation as they focus on single aspects in RE, such as problems in RE processes or RE improvements, or 
they focus on specific countries. Thus, these studies by now remain not representative.

To close this gap in literature, we designed a family of surveys in joint collaboration with different researchers. The design 
of the survey as well as the interpretation of the results are both conducted along an initial theory. 
By relying on an initial theory built on the basis of or experiences and available studies, and by bringing together 
different interdisciplinary communities during replications, the family of surveys shall contribute to an empirical 
basis for empirical generalisations and problem-driven research in RE.

\subsection{Previously Published Material} 
\label{sec:PreviouslyPublishedMaterial}

We previously published the basic survey design and discussed preliminary results from Germany in conference 
proceedings~\cite{MW2013a}. Those results were obtained from an intermediate data export and contained answers 
from 30 respondents (see also Fig.~\ref{fig.SurveyDesign} on page \pageref{fig.SurveyDesign}). The full data set is published in addition as a technical report~\cite{MW13b}, which we used to inform the participants about the results. We interpreted those results with basic descriptive statistics and reasoning.

In the article at hand, we rely on the full results from the completed survey including answers from 73 companies of which we use 58 as those responses are complete. We extend our previously published intermediate analysis with a full data analysis including:
\begin{compactenum}
\item A detailed theory with a series of hypotheses we inferred from our own experiences and observations as well as from available literature
\item A detailed data analysis including statistical analysis and the application of Grounded Theory to comprehensively analyse quantitative data as well as qualitative data
\end{compactenum}

\section{Theory and Expectations}
\label{sec:Theory}

Before we go into the design of the survey in the next section, we describe the theory and expectations we have about the
state-of-practice and problems in requirements engineering which we will test using the survey results. The basis for this 
theory is mostly our observations and experiences in industrial collaborations as well as our (incomplete) impression of the
existing literature. For future work, a possibility would be to run a systematic literature review as a more complete foundation
of the theory. Our intention is that using the family of surveys that runs continuously, we can iteratively refine and extend
the theory using the survey results and interactions between the participating researchers.

We use the term \emph{theory} in the sense of a social theory~\cite{Popper02, AHK11} and refer to a set of falsifiable and 
testable hypotheses having an inductive nature (see also Tab.~\ref{tab:terminology}). The underlying objective of our research
is to improve requirements engineering in practice. The study shall serve the purpose of getting a better empirical understanding 
on the status quo in requirements engineering as well as contemporary problems so that we can infer common improvement goals. 

Hence, we structure our initial social theory into four subsets of hypotheses. We first need to describe the current status of requirements 
engineering practice. Second, similarly, we describe hypotheses on the status quo in requirements engineering improvement. As the 
status quo might not reflect what practitioners expect from a good RE, we want to understand the expectations practitioners have on 
good requirements engineering practice. We focus in particular on expectations on company-specific RE standards. Lastly, to
derive improvement goals, we formulate our theory about concrete problems in RE practice.


In the following, we define the concrete hypotheses for each theory subset. Each hypothesis is intended to be falsifiable and
testable by statistical analysis later based on the survey results. We state all hypothesis in the alternate form. The null hypothesis is always that the distribution of the answers is symmetric about a central value $\mu$. In our theory, we describe our alternative hypotheses that the true location is greater or less than that, respectively. For easier tracing of the theory to the questionnaire used in the survey, we 
provide the question number as well as our alternative hypothesis. In case of answer 
sets using a Likert scale, the alternative hypothesis is either $>3$ meaning that the median rating was \emph{important}/\emph{very important} or \emph{agree}/\emph{strongly agree}
or $<3$ meaning that the median rating was \emph{unimportant}/\emph{very unimportant} or \emph{disagree}/\emph{strongly disagree}. For easier readability of the
table, we denote $>3$ as ``$+$'' and $<3$ as ``$-$''.

In case of yes/no questions,  the numerical value is $>0$ for ``yes'' (0 would be ``no''). To ease the readability, we group different hypotheses for all 
answer possibilities to one single hypothesis if possible. 

\subsection{Expectations on Good Requirements Engineering} 

Before we describe our theory on the actual practices and problems in RE, we first describe our understanding of the expectations 
of practitioners how good requirements engineering should look like. All hypotheses in this theory subset are shown in 
Tab.~\ref{tab:hypothesisRQ1}. As a prerequisite, we assume that practitioners consider improvement in requirements engineering
beneficial but also challenging. Yet, we also expect that overall practitioners will consider improvements in all development phases
and areas, like project management or implementation, beneficial but challenging. We expect RE improvement to be challenging due 
to the volatility in the process~\cite{MLPW11}, but also, for example, quality assurance due to the difficulty of developing and applying 
adequate (valid) metrics and measurements (see, e.g.,~\cite{DOJC+93}). 

\begin{table}[!htb]
\centering \scriptsize
\caption{Hypotheses about the expectations on good requirements engineering.}
\label{tab:hypothesisRQ1}
\begin{tabular}{lp{0.7\linewidth}c}
\hline
\textbf{No.} & \textbf{Hypotheses on improvement (Q~9 and Q~10)} & \textbf{Expec.} \\ \hline 
H~1 & The improvement of all development phases is beneficial. &  $+$ \\ 
H~2 & The improvement of all development phases is challenging.      &  $+$ \\ \hline
& \textbf{Hypotheses on general expectations on standardisation in RE (Q~11)} &  \\\hline
H~3 & The standardisation of RE does not hamper creativity.      &  $-$ \\ 
H~4 & The standardisation of RE improves the overall process quality.    &   $+$ \\ 
H~5 & Standardised document templates and tool support benefits the communication. & $+$ \\ 
H~6 & The structure of documents should be standardised across different project environments, but the process itself should be left open for project participants. & $+$ \\ \hline
& \textbf{Hypotheses on particular aspects in company-specific RE standards (Q~12)} & \\\hline
H~7 & The definition of standard methods and modelling techniques is not important. & $-$ \\
H~8 & Tool support for V\&V of req. spec. is not important. & $-$ \\ 
H~9 & The definition of standardised RE artefacts with document templates and/or tool support is important. & $+$ \\
H~10 & Tailoring mechanisms according to project characteristics are important. & $+$ \\ 
H~11 & The definition of roles and responsibilities is important. & $+$ \\ 
H~12 & Support of impact analysis is important.  &  $+$ \\ 
H~13 & Deep integration with other disciplines is important.     &  $+$ \\ 
 & Support of agility in the process is\ldots  &  \\
H~14 &     \ldots important in case of business information systems. & $+$ \\ 
H~15 &     \ldots not important in case of embedded systems. & $-$ \\
& Support of prototyping is\ldots    &  \\
H~16 &     \ldots important in case of embedded systems. & $+$ \\ 
H~17 &     \ldots not important in case of business information systems. & $-$ \\ \hline
& \textbf{Hypotheses on reasons for defining a company-specific RE standard (Q~13)} & \\\hline
H~18 & Formal prerequisites for project acquisition do not motivate a standard.       &  $-$ \\
H~19 & Support of progress control does not motivate a standard.       &  $-$ \\ 
H~20 & Seamless development by integrating RE into the development process motivates a standard.      &  $+$ \\
H~21 & Better tool support motivates a standard. & $+$ \\
H~22 & Support of distributed development motivates a standard.       &  $+$ \\
H~23 & Better quality assurance of artefacts motivates a standard.      &  $+$ \\
H~24 & Support of benchmarks motivates a standard.  &  $+$ \\
H~25 & Support of project management and planning motivates a standard.     &  $+$ \\
H~26 & Higher efficiency motivates a standard.   &  $+$ \\
H~27 & Knowledge transfer motivates a standard.     &  $+$ \\ 
 & Compliance to regulations and standards (like CMMI)\ldots      &  \\
H~28 &     \ldots does not motivate a standard in case of small/medium-sized companies. & $-$ \\ 
H~29 &     \ldots motivates a standard in case of large companies (more than 2000 employees). & $+$ \\\hline
& \textbf{Hypotheses on barriers for defining a company-specific RE standard (Q~14)} & \\\hline
H~30 &  Higher demand for communication does not barrier defining a standard.  &  $-$\\  
H~31 & Higher process complexity barriers defining a standard.  & $+$\\
H~32 & Lower efficiency barriers defining a standard. &  $+$\\
H~33 & Missing willingness to change barriers defining a standard. & $+$\\
H~34 & Missing possibilities of standardisation barrier defining a standard. &  $+$\\ \hline
\end{tabular}
\end{table}

The subsequent set of hypotheses initially characterises the expectations practitioners have on a good RE. We concentrate on hypotheses
about the RE standards that are or should be in place in their companies. Due to the known criticality and importance of RE~\cite{broy06_mbRE}, 
its sensitivity to the customer domain, and the underspecified terminology, e.g., in the area of non-functional requirements~\cite{Glinz05}, we expect 
the practitioners to rate a standardisation of RE as increasing the overall quality. The terminological problems in RE and the consequences of the 
process complexity on the quality of the artefacts created are also reasons why we believe the standardisation of artefacts is important~\cite{MWLBC10}.

After the general expectations on RE standardisation, we define hypotheses according to our experiences about the expectations 
practitioners have on particular aspects of their own standards. For instance, based on our investigation published in~\citep{MLPW11, MPKB10}, we expect 
practitioners to demand standards that focus on the RE artefacts, clear roles and responsibilities, and tailoring mechanisms rather than on 
strict processes and methods to allow for more flexibility and a better communication. 

While we see tool support to be a project-specific decision that often cannot be dictated in multi-project environments, we believe aspects of 
process integration (e.g., with change management) to be rated very important~\cite{GW06}. Regarding the support for agility and the application 
of prototyping, we expect it to be rated as important while practitioners working in the embedded systems domain should have limited possibilities 
for agility because of different suppliers with restricted access to the customers domains and the difficulty of combining hardware/software co-design
in agile approaches. Our expectation is also that the embedded systems domain will see the use of prototypes as important, e.g., due to the 
challenges arising from software/hardware co-design. 

Finally, we define hypotheses for the motivation and barriers practitioners expect when defining an RE standard. Based on similar 
observations as in the previous hypotheses, we expect practitioners to see the improvement of the quality in the RE artefacts to be the main 
motivation for defining a company-specific RE standard. This should especially hold for small/medium-sized companies which do not see compliance to 
regulations and (improvement) standards as a motivation due to their general reluctance to such standards~\cite{SNJABR07}. Relying, for example, 
on the observations made by Damian et al.~\cite{DC06}, we also suppose practitioners to agree on the support of knowledge transfer and, thus, on 
the need of defining artefact models in the company standards, because artefact models make implicit knowledge about the domain explicit (e.g., 
with templates or modelling guidelines)~\cite{MPKB10}. As a barrier to defining an RE standard, we rely on our experiences in research cooperations 
and expect practitioners to agree on a higher process complexity, the missing willingness for change and the missing possibilities for standardisation.

\subsection{Status Quo in Requirements Engineering} 

Following the establishment of hypotheses on RE improvement and the expectations of practitioners on RE standardisation, we focus on the
current state of RE in practice in this second subset of our theory. We describe the RE practice in the areas of elicitation, process tailoring,
change management, and the used RE standard. The hypotheses are shown in Tab.~\ref{tab:hypothesisRQ2}. 

\begin{table}[!htb]
\centering \scriptsize
\caption{Hypotheses on the status quo in requirements engineering practice.}
\label{tab:hypothesisRQ2}
\begin{tabular}{lp{0.7\linewidth}c}
\hline
\textbf{No.} & \textbf{Hypotheses on the elicitation of requirements (Q~16)} & \textbf{Expec.} \\\hline
H~35 & Requirements are elicited via workshops. & yes \\ 
H~36 & Requirements are elicited via change requests. & yes \\ \hline
& \textbf{Hypotheses on defining and tailoring an RE standard (Q~18 and Q~22)} & \\\hline
H~37 & Requirements engineering standards are defined due to company-specific demands. & yes \\ 
H~38 & The RE standard is tailored at the beginning of a project by the project lead based on experiences. & yes \\ \hline
& \textbf{Hypotheses on the characteristics of their RE standard (Q~19)} & \\\hline
H~39 & The standard does not rely on an architecture model with different levels of abstraction & $-$ \\
H~40 & The standard does not include a differentiated view on different requirements classes and dependencies. & $-$ \\ 
H~41 & The standard does not includes tracing relationships among the contents. & $-$ \\
H~42 & The standard does not include a differentiated view on non-functional requirements. & $-$ \\ 
H~43 & The standard includes a differentiated view on different requirements classes, but no dependencies. & $+$ \\ \hline
& \textbf{Hypotheses on the application of the RE standard (Q~21 and Q~23)} & \\\hline
H~44 & Each project can decide whether to use the standard. & yes \\ 
H~45 & The application of the RE standard is controlled via analytical quality assurance. & yes \\ \hline
& \textbf{Hypothesis on change management (Q~20)} & \\\hline
H~46 & A requirements change management is established after formally accepting a requirements specification. & yes \\ 
\hline
\end{tabular}
\end{table}

Regarding the elicitation of requirements, we expect practitioners, especially in large companies, to conduct workshops and otherwise 
rely on formal change management via change requests, because we experience German companies not yet to be affine to agile methods.

Instead of factors lying outside the company, we expect that RE standards are defined because of company-specific demands. Based
on our experiences, we also expect that the RE standard is tailored by the project lead based on his or her experience at the beginning
of each project instead of using a defined tailoring approach. The main reason is that tailoring itself is a topic that is still not well 
explored~\cite{PPLB07}. This especially holds for RE~\cite{Mendez11}. 

We generally expect the company-specific RE standards to be immature compared to other disciplines due to the inherently complex 
nature of RE. We rely, for example, on the observations of \cite{HBR02} and suppose the standards to define coarse processes rather 
than well defined artefact models that support traceability~\cite{RJ01}. Also, we expect the standards to distinguish different requirements 
classes but not to include a detailed view with clear (tracing) dependencies. This is especially true for non-functional aspects as quality 
itself is a multifaceted topic with various domain-dependent interpretations~\cite{Glinz05, 1984_garvind_product_quality, KP96}.

In consequence, we expect the application of their standards not to be mandatory. Still, given multi-staged bidding procedures and complex, 
formal tendering procedures as well as off-shoring being important topics for German companies, we expect the requirements specifications 
to be controlled as part of an analytical quality assurance, e.g., as part of a quality gate (see, e.g.,~\cite{SEH09}).

Finally, for the state of practice in change management, we expect that it is still the case that a requirements change management is
established after a requirements specification (expected to be complete) was formally accepted. This is due to our observations that
agile approaches for handling requirements change are not widespread in Germany yet.

\subsection{Status Quo in Requirements Engineering Improvement} 

Similar to the theory subset before, we look into the current state of practice; this time the practice in RE improvement. We see process
improvement important for any software engineering discipline but for the volatile and complex RE, we expect this to be essential.
Our hypotheses are shown in Tab.~\ref{tab:hypothesisRQ3}. 

\begin{table}[!htb]
\centering \scriptsize
\caption{Hypotheses on requirements engineering improvement.}
\label{tab:hypothesisRQ3}
\begin{tabular}{lp{0.7\linewidth}c}
\hline
\textbf{No.} & \textbf{Hypotheses on continuous improvement (Q~24 and Q~25)} & \textbf{Expec.} \\ \hline 
H~47 & Requirements engineering is not continuously improved. & no \\ 
H~48 & A continuous improvement is done to determine strengths and weaknesses. & yes \\ \hline
& \textbf{Hypotheses on RE improvement (Q~26, Q~27, and Q~29)} & \\\hline
& RE improvement is done systematically\ldots  & \\
H~49 & \ldots via an own business unit / role in case of large companies (more than 2000 employees). & yes\\ 
H~50 & \ldots via external consultants in case of large companies (more than 2000 employees). & yes \\ 
H~51 & \ldots by project participants in case of small/medium-sized companies. & yes\\ 
H~52 & RE improvement is done via internally defined standards/approaches. & yes \\ 
H~53 & RE assessments/audits are done using qualitative methods. & yes \\ \hline

\end{tabular}
\end{table}

Given the plethora of different (not necessarily compatible) methods and practices for RE and the still low number of reported evidence 
on applying those methods in practice~\cite{CDW12}, as well as the inherent complexity and volatility of RE itself, improving RE (going 
beyond audits and assessments) remains a difficult topic with only few known concepts~\cite{PIGO08, MW2013}. For this reason, we 
believe that most participants do not improve their RE continuously. For those who do, we expect the main reason to be able to determine
strengths and weaknesses of their RE process and standard.

For those that improve their RE, we rely on the observations made by Staples et al.~\cite{SNJABR07} in the area of software process 
improvement. Large companies use assessments conducted by an own business unit or via external consultants, 
e.g., as part of a certification program (see also~\cite{NMJ09}). Small and medium-sized companies, however, do not have the resources
to have separate improvement groups. Instead, they will conduct improvements by project participants directly.
We believe that normative improvement approaches are losing industrial attention as they are steered by externally defined 
goals and norms with the purpose of assessing and benchmarking different companies. We consequently expect that qualitative RE 
improvement approaches are predominantly used~\cite{PIGO08, MW2013} based on internally defined standards and approaches.  

\subsection{Contemporary Problems in Requirements Engineering}

Finally, the last theory subset describes our main area of interest of our study: contemporary problems of practitioners in their RE
practice. We define hypotheses about problems in the RE standard and their projects in Tab.~\ref{tab:hypothesisRQ4}.

\begin{table}[!htb]
\centering \scriptsize
\caption{Hypotheses on contemporary problems in requirements engineering.}
\label{tab:hypothesisRQ4}
\begin{tabular}{lp{0.7\linewidth}c}
\hline
\textbf{No.} & \textbf{Hypotheses on problems with the RE standard (Q~31)} & \textbf{Expec.}\\\hline
H~54 & It is not too hard to understand. & $-$ \\ 
H~55 & It is not too complex. & $-$ \\ 
H~56 & It scales to the projects' high complexity. & $-$ \\ 
H~57 & It sufficiently defines roles and responsibilities. & $-$ \\ 
H~58 & It is too abstract. & $+$ \\ 
H~59 & It does not support the specification of precise requirements. & $+$ \\ 
H~60 & It is too heavy weight for our projects. & $+$ \\ 
H~61 & It is not flexible enough. & $+$ \\ 
H~62 & It does not sufficiently define a clear terminology. & $+$ \\ 
H~63 & It gives no guidance on how to create the specification documents. & $+$ \\ 
H~64 & It does not sufficiently allow for deviations according to project circumstances. & $+$ \\ 
H~65 & It is not sufficiently integrated into project management. & $+$ \\ 
H~66 & It is not sufficiently integrated into design and architecture. & $+$ \\ 
H~67 & It is not sufficiently integrated into risk management. & $+$ \\ 
H~68 & It is not sufficiently integrated into test management. & $+$ \\ \hline
& \textbf{Hypotheses on RE problems within projects (Q~32)} & \\\hline
H~69 & Communication flaws within the project team are not a problem. & $-$ \\ 
H~70 & Insufficient support by the project lead is not a problem. & $-$ \\ 
H~71 & Gold plating (implementing features without corresponding requirements) is not a problem. & $-$ \\ 
H~72 & Weak knowledge of the customer's application domain is not a problem. & $-$ \\ 
H~73 & Technically unfeasible requirements are not a problem. & $-$ \\ 
H~74 & A weak relationship to the customer is not a problem. & $-$ \\ 
H~75 & A volatile customer business domain (e.g., changing business requirements) is not a problem. & $-$ \\ 
H~76 & Communication flaws between project team and customer are a problem. & $+$ \\ 
H~77 & There are terminological problems. & $+$ \\ 
H~78 & Unclear responsibilities are a problem. & $+$ \\ 
H~79 & Incomplete and/or hidden requirements are a problem. & $+$ \\ 
H~80 & Insufficient support by the customer is a problem. & $+$ \\ 
H~81 & Stakeholders with difficulties in separating requirements from previous solutions are a problem. & $+$ \\ 
H~82 & Inconsistent requirements are a problem. & $+$ \\ 
H~83 & Missing traceability is a problem. & $+$ \\ 
H~84 & Moving targets are a problem. & $+$ \\ 
H~85 & Weak access to customer needs and/or business information is  a problem. & $+$ \\ 
H~86 & Time boxing and not enough time in general is a problem. & $+$ \\ 
H~87 & The discrepancy between the high degree of innovation and the need of formal requirements acceptance is a problem. & $+$ \\ 
H~88 & Underspecified requirements that are too abstract and allow for various interpretations are a problem. & $+$ \\ 
H~89 & Unclear and unmeasurable non-functional requirements are a problem. & $+$ \\ \hline
\end{tabular}
\end{table}

Regarding the problems in the RE standards used by practitioners, we again make use of experiences we have made. That 
is, we take into account that an RE standard can either follow the activity-based paradigm, where coarse activities, methods, 
and practices are defined, or the artefact-based paradigm, where the standard is defined on the basis of a process-agnostic 
artefact model that defines which artefacts to produce and which relationships to take into account (see also~\cite{MPKB10}). 
Our experiences on the establishment of an RE standard and the implications the choice of a particular paradigm has at project 
level results from experiments, case studies, and field studies we conducted in this area~\cite{KMM2013, MLPW11, MWLBC10}. 
Based on those empirical investigations and the fact that most standards follow the activity-based paradigm as artefact orientation is still a young philosophy with various interpretations and manifestations in practice~\cite{Mendez11}, we formulate corresponding expectations on problems practitioners see in their standards. 

We expect the standards to insufficiently define terminology and to not guide the creation of syntactically complete and consistent 
artefacts as the focus lies on potentially incompatible methods and modelling techniques rather than on clearly defined artefacts, 
modelling concepts, and their dependencies.  Whereas we expect the standards to be seen as easy to understand and consequently 
to be seen as too abstract, we expect practitioners to rate the standard to be too heavy-weight for their projects. Furthermore, as 
the integration of RE into different phases of the software process life cycle is still an open research area, we believe the practitioners 
to have problems with this integration.

Regarding more general RE problems in practical projects, we take into account the broad spectrum of empirical investigations introduced 
in Sect.~\ref{sec:RelatedWork} including the investigations of Hall et al.~\cite{HBR02} and Hsia et al.~\cite{HDK93} on problems in RE in 
practical settings as well as investigations on factors that generally influence the performance of RE at project level~\cite{MWLBC10}.

According to those investigations, we formulate that practitioners have problems with inconsistent requirements as (beside other reasons) 
activity-oriented standards do not guide the creation of precise results~\cite{MLPW11}. Another example is the problem of having weak 
access to customer needs, because this was stated as a major project influence that needs to be taken into account in a previous 
investigation~\cite{MWLBC10}. In the end, we also expect companies that are not aware of their RE artefacts to state to have more 
problems with the completeness and consistency in the project-specific specification documents.

\section{Design of a Family of Surveys}
\label{sec:Design}

In the following, we introduce the design of the family of surveys. Our objective is to herewith lay the foundation for a survey to be continuously replicated in different countries. With those replications, we aim at generalising the state of practice and trends in RE including practitioners' expectations, the status quo in RE and its improvement, and contemporary problems experienced in companies. To support the dissemination of the results and the collaboration among the research communities, we provide for each replication round a shared survey infrastructure using the same questionnaire, and we disclose the anonymised data to the PROMISE repository (see also the project website \url{RE-Survey.org} for further information).

In the following, we formulate five research questions that build the frame for our surveys. Afterwards, we design the overall methodology and introduce the instrument, i.e.\ the (typed) questions and the categories, in Sect.~\ref{sec:Instrument}. In Sect.~\ref{sec:DataAnalysis}, we present the data analysis procedures, before concluding with the presentation of the validity procedures in Sect.~\ref{sec:ValidityProcedures}.

\subsection{Research Questions}
We formulate five research questions, shown in Tab.~\ref{tab:rqs}, to steer the overall design of the surveys.

\begin{table}[htb]
\caption{Research questions.\label{tab:rqs}}
\begin{tabular}{p{0.13\linewidth}p{0.81\linewidth}}
\hline
\textbf{RQ~1} & What are the expectations on a good RE ? \\ 
\textbf{RQ~2} & How is RE defined, applied, and controlled? \\
\textbf{RQ~3} & How is RE continuously improved? \\
\textbf{RQ~4} & Which contemporary problems exist in RE, and what implications do they have?\\
\textbf{RQ~5} & Are there observable patterns of expectations, status quo, and problems in RE?\\
\hline
\end{tabular}
\end{table}

Not included in those research questions (but in the instrument) are questions to characterise the survey respondents and the industrial environment in which they are involved. The first research question aims at investigating the expectations and preferences the respondents have on a good RE. Research question~2 and~3 have the goal of investigating the status quo in the RE as it is established in their companies as well as industrial undertakings to continuously improve RE. Research question~4 aim at investigating which problems practitioners experience in their project environments, what implications these problems have on the product and process, and to what extent those problems have already lead to failed projects. Finally, research question~5 considers the analysis of statistically significant correlations in the result set of RQ~1-4 including the variables used to characterise the study population.

\subsection{Methodology}
\label{sec:Methodology}

\begin{figure*}[!hbt]
\centering
  \includegraphics[width=0.7\textwidth]{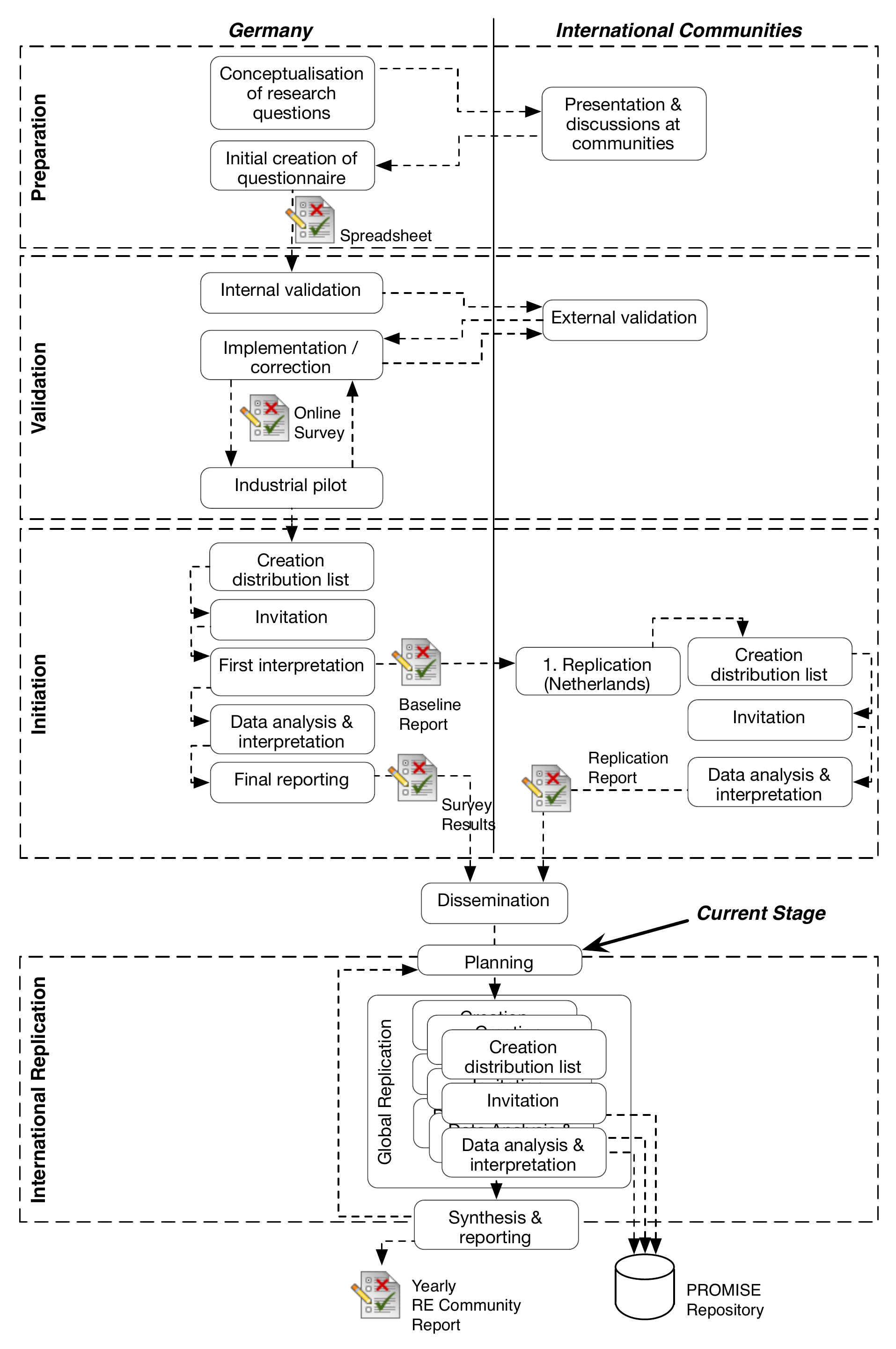}\\
  \caption{Overview of the methodology.}\label{fig.SurveyDesign}
\end{figure*}

We designed our family of surveys based so that we can test the theory defined in Sect.~\ref{sec:Theory} and find new hypotheses to extend the theory. This design contains four stages, which we illustrate in a simplified manner in Fig.~\ref{fig.SurveyDesign}. We distinguish between activities performed in isolation in Germany and activities where we actively involved, or will involve again, international research communities. 

The first two stages contain the activities carried out to design and validate the survey structure (the research questions and the instrument). The third stage consists of the survey implementation and the initial start in Germany from which we drew a baseline to report our intermediate findings in~\cite{MW2013a}. The last stage contains the survey replications which began at the time of writing this article. 

Considering the notion of ``replication'', we rely on the classification introduced by G\'{o}mez et al.~\cite{GJV10} and aim at empirical generalisations. Each replication of the survey is performed  independently by different researchers in different countries using the same infrastructure and instrument, before synthesising and reporting the overall results in joint collaboration. For the initial start of the survey, we rely on the described theory for defining the research questions and the instrument. Due to an expected learning curve on basis of the results, we expect that the theory will become more mature and change from year to year affecting the variables in the instrument.

In the following, we introduce the four stages of our methodology in more detail. The resulting instrument of the survey is introduced in the next Sect.~\ref{sec:Instrument}.

\subsubsection{Preparation}

Based on discussions at international events and the experiences we made during previous studies, we conceptualised an initial set of research questions and jointly discussed them at different community forums. The background and the thematic frame for the research questions was \emph{investigating the status quo in RE and its improvement in industry} as well as \emph{contemporary problems practitioners have in their professional project setting} to reason for improvement goals.  

We presented the idea of a joint survey at several international workshops and conferences as well as in personal communications with various researchers in requirements engineering and empirical software engineering around the world.
After checking for the resonance and agreeing with other researchers on a joint collaboration in the envisioned topic, we created an initial version of the instrument to answer our research questions. This questionnaire includes, where possible and reasonable, closed questions for a clear data analysis and to keep the effort low for practitioners when answering the questionnaire. To maximise the validity, we performed a series of validation tasks, which we introduce in the following.

\subsubsection{Validation}
\label{sec:Validation}
After creating the initial questionnaire, we performed a series of validation tasks, which took us in total three months. We first performed an internal validation of the questionnaire with a review by researchers not involved in the design of the questionnaire at the Technische Universit\"at M\"unchen and the University of Stuttgart. This internal validation should ensure that the closed questions are clearly interpretable and sufficiently complete to answer the research questions, i.e., it should increase the internal and the construct validity. For the external review, we invited several researchers from different universities of which the ones listed in Tab.~\ref{tab:review} could do the review in the short time frame we could allow. 

\begin{table}[htb]
\caption{Involved researchers.
\label{tab:review}}
\begin{tabular}{p{0.2\linewidth}p{0.7\linewidth}}
\hline
\textbf{Review} & \textbf{Researcher}  \\
\hline 
Internal review & M. Broy, S. Eder, J. Eckhardt, K. Lochmann, J. Mund,  B. Penzenstadler \\ 
External review & M. Daneva, R. Wieringa (Twente)\\
& M. Genero (Castilla-La Mancha) \\
& J. M\"unch (Helsinki)\\
\hline
\end{tabular}
\end{table}

After increasing the construct validity with the external validation, we implemented the survey as a Web application using the \emph{Enterprise Feedback Suite}. With this web-based version, we conducted an industrial pilot phase with an industry participant. This participant has worked for five years as process consultant and has deep insights into the envisioned application domains, used RE standards, and he is familiar with the terminology used. His feedback and the analysis of the responses served to identify vague questions, incomplete answers in the closed questions, and how those answers apply to his context, thus, increasing the internal and the external validity. 
We complemented this pilot with two additional dry runs and external validations, before re-setting the data tables for the initiation of the survey.

\subsubsection{Initiation}


The initiation phase contains mostly the survey conducted in Germany. The survey was closed and by invitation only to allow for a transparent and reproducible response rate and to ensure that the survey was answered by not more than one representative contact person per company (or business unit in case of large enterprises). In addition, the survey is anonymous due to the criticality of the questions. This paper includes, besides this design of the family of survey, the full data analysis form the German survey. The survey in Germany was online from November 17th, 2012, until January, 31st, 2013. 

In each survey round, we begin with the creation of a distribution list. This distribution list contains research partners from the universities while aiming at different roles from different companies of different sizes and application domains. Where possible, we inform the partners in advance and select, where reasonable, appropriate contact persons to support high response rates. The official invitation contains
\begin{compactitem}
\item the basic information about the goals of the survey, the categories of questions and the context of the survey as part of a global family of surveys, and
\item the link to the survey and a password.
\end{compactitem}

We additionally ask the participants to forward the invitation if necessary and to inform us about the number of participants to support the inference of the response rate. We further re-ensure the participants about the anonymous nature of the survey and that they can add their e-mail address at the end of the questionnaire (not associated with the answers) so that we can inform them about the final results as an incentive. 

We triggered a first replication in the Netherlands after drawing the first results we concluded from a baseline report -- previously presented in~\cite{MW2013a}. The results as well as experiences from this first replication will serve to further disseminate the survey among the different RE-related and empirical research communities as well as improvements to the design and theory. 

To this end, we use the current theory and questionnaire and modify and extend it with the results presented in this article. That is, we discuss given questions based, e.g., on whether we could confirm corresponding hypotheses from our theory and we change the questionnaire accordingly (see also the the discussions of our results in relation to our theory in Sect.~\ref{sec:Results} and the lessons learnt discussion in Sect.~\ref{LessonsLearnt}).

\subsubsection{International Replications}

We will hold calls and workshops for the detailed planning of the next iterations with the researchers from different countries with whom we have already involved or agreed to join the endeavour. We plan each replication round to be performed in isolation by different researchers using the same pre-agreed theory, corresponding questionnaire, and the survey infrastructure provided by us. The replications are performed, after a planning phase, independently and the survey design will change over the years due to a certain learning curve. 

To ensure a reproducible generalisation and the openness of the results to the communities, we will disclose the anonymised results to the PROMISE repository\footnote{\url{ http://promisedata.googlecode.com}}. The overall aim is to establish a representative and open data basis to investigate industrial trends in RE. In addition, we will invite all replications from each round to join us in synthesising all the results and data analysis to discuss global insights and trends as well as differences between different countries.


\subsection{Survey Instrument}
\label{sec:Instrument}

Table~\ref{tab:instrument} summarises the questions of the survey in a simplified and condensed manner. We define in total 35 questions grouped according to the research questions, except for the fifth research question that considers a correlation analysis among all answers given. We begin with a set of questions to characterise the respondents and the companies in which they work. At the end of the survey, the respondents can enter their e-mail address and freely add any other aspect that remained unaddressed in the survey (used as input for future modifications of the instrument).
\begin{table*}[tb]
\centering \scriptsize
\caption{Questions (simplified and condensed).}
\label{tab:instrument}
\begin{tabular}{crp{0.65\linewidth}l}
\hline
\textbf{RQ}& \textbf{No.}& \textbf{Question} & \textbf{Type} \\ \hline 
--           & Q~1        & What is the size of the enterprise? & Closed(SC)\\
            & Q~2           & What is the main business area of your company? & Closed(MC)\\
             & Q~3          & Does your company participate in globally distributed projects? & Closed(SC)\\
             & Q~4          &  In which country are you personally located? & Open \\
             & Q~5          &  In which application domain/branch are you most frequently involved in your projects? & Closed(MC)\\
            & Q~6           & To which project role are you most frequently assigned to in those projects? & Closed(SC) \\
            & Q~7           &  How would you classify your experience as part of this role? & Closed(SC)\\
            & Q~8           &  Which organisational role takes your company usually in aforementioned projects? & Closed(SC)\\ \hline
RQ~1  & Q~9            &  How beneficial would you rate an improvement for following disciplines in your company?	 &  Likert  \\
            & Q~10           &  How challenging would you rate an improvement for following disciplines in your company?	 & Likert  \\
            & Q~11           &  Please rate the following statements on RE standardisation according to your expectations.			       & Likert  \\
            & Q~12           &  How important would you consider the following aspects when defining an RE standard?	& Likert  \\
             & Q~13          &  Which reasons do you agree with as a motivation to define an RE standard?	& Likert \\
            & Q~14           &  Which reasons do you see as a barrier to define an RE standard? & Likert \\ \hline
RQ~2   & Q~15           &  How would you classify you/your company to be involved in RE? & Closed(SC)\\
            & Q~16           &  If you elicit requirements in your regular projects, how do you elicit them? & Closed(MC) \\
            & Q~17           &  What RE standard have you established at your company? & Closed(MC) \\
            & Q~18           &  Which of the following reasons apply to the definition of an RE standard in your company?	 & Closed(MC) \\
             & Q~19          &  How would you rate the following statements to apply to your RE standard?				 & Likert  \\
            & Q~20           &  How is your change management defined regarding your RE? & Closed(MC)\\
            & Q~21           &  Which of the following statements apply to the project-specific application of your RE standard? & Closed(MC)\\
            & Q~22           &  How is your RE standard applied (tailored) in your regular projects?	 & Closed(MC) \\
            & Q~23           &  How is the application of your RE standard controlled? & Closed(MC) \\ \hline
RQ~3  & Q~24            & Is your RE continuously improved? & Closed(SC) \\
            & Q~25           &  What would you consider to be the motivation for a continuous improvement? & Closed(MC)\\
            & Q~26           &  Which of the following statements applies regarding the continuous RE improvement?& Closed(MC)\\
             & Q~27          &  Do you use a normative, external standard for your improvement? & Closed(SC) \\
            & Q~28           & If you use an internal improvement standard and not an external one, what where the reasons? & Open\\
            & Q~29           & Which methods do you use for your RE improvement (regarding assessments/audits)? & Closed(MC)\\
            & Q~30           & If you use metrics to assess your RE in the projects, which ones would you deem most important?& Open \\ \hline
RQ~4  & Q~31            &  Please rate the following statements for your RE standard according to your experiences. & Likert  \\
              & Q~32         &  How do the following (more general) problems in RE apply to your projects? & Likert \\
            & Q~33           &  Considering your personally experienced problems (stated in the previous question), which ones would you classify as the five most critical ones (ordered by their relevance)? & Closed \\
                & Q~34       &  Considering your personally experienced most critical problems (selected in the previous question), how do these problems manifest themselves in the process, e.g., in requests for changes? & Open \\
               & Q~35        &  Considering your personally experienced most critical problems (selected in the previous question), which would you classify as a major cause for project failures (if at all)? & Closed(MC) \\ \hline
\end{tabular}
\end{table*}

For each question in the table, we denote whether it is an open question or a closed one and whether the answers are mutually exclusive single choice answers (SC) or multiple choice ones (MC). Most of the closed multiple choice questions include a free text option, e.g., ``other'' so that the respondents can express company-specific deviations from standards we ask for. We furthermore use Likert scales on an ordinal scale of 5 and defined for each a maximum value (e.g., ``agree'', or ``very important''), a minimum value (e.g., ``disagree'', or ``very unimportant''), and the middle (``neutral''). The latter allows the respondents to make a selection when they have, for example, no opinion on the given answer options. Finally, we define Q~17 and Q~24 as conditional to guide through the survey by filtering subsequent question selection. For instance, if respondents state in RQ~17 that they have not defined any company-specific RE standard, the last questions of this section are omitted. 

The full questionnaire (including which expected answers were confirmed and which ones were rejected) is available in the publication section of the website~\url{RE-Survey.org}. 
\subsection{Data Analysis}
\label{sec:DataAnalysis}

The data of the survey contains a mix of information about the companies and the RE standards used and expert opinions of the subjects involved in those companies. Moreover, each survey (our base and further replications) does not rely on random samples as we opt for industry participants to whom we have contact, even if the participants distribute the invitation to further colleagues. Finally, regarding the expert opinions, we express the subjects' opinions with Likert scales which are specified with ordinal scales with no interval data, i.e., the distances between the single values in the variables (e.g., ``strongly agree'', ``agree'', and ``disagree'') are not equally distributed. In other cases, we have open questions or certain variables in nominal scales, e.g., the companies either apply certain methods for their RE improvement or they do not. For this reason, we need to rely on different procedures for the data analysis. We will introduce those procedures structured according to the research questions in the following.

\paragraph{Statistical Analysis for RQ~1--3}
For the analysis of the answers to RQ~1--3, we first rely on descriptive statistics to get a better understanding of the data. We use the mode and median for the central tendency of the ordinal data and the median absolute deviations (MAD) for its dispersion. For the nominal data, we calculate the share of respondents choosing the respective option. 

Second, we apply the one-sided, unpaired Wilcoxon signed rank test to test our theories discuss in detail in Sect.~\ref{sec:Theory}. 
We test the hypotheses at a confidence level of 0.95 and we will present the resulting p-values with an accuracy of four digits after the decimal point.

\paragraph{Descriptive Interpretation and Grounded Theory for RQ~4}
To answer RQ~4, i.e., the analysis of contemporary problems in RE, we first quantify the answers given for the selection of the predefined problems the participants shall rank as they have experienced them in their projects. As part of this quantification, we also sum up to which extent the given problems have led to project failures in the opinion of the participants.

For analysing the answers given to the open question on what implications the RE problems have (Q~34), we apply the Grounded Theory method. In contrast to the previously applied analysis procedures where we confirm or reject the hypotheses defined in our theory (see Sect.~\ref{sec:Theory}), the answers given for RQ~4 are not related to any expectations yet whereby we follow an inductive logic approach to generate a new theory based on given qualitative data, respectively to extend our previously defined theory. Grounded Theory, in its essence~\cite{GS67, AHK11}, considers the four following basic steps in an inductive bottom-up approach: 
\begin{compactenum}
\item \emph{Open coding} to analyse the data by adding codes (representing key characteristics) to small coherent units in the answers, and categorising the developed concepts in a hierarchy of categories as an abstraction of a set of codes -- all repeatedly performed until reaching a state of saturation. We define the (theoretical) saturation as the point where no new codes (or categories) are identified and the theory is convincing to all participating researchers~\cite{BM11}.
\item \emph{Axial coding} to define relationships between the concepts, e.g., ``causal conditions'' or ``consequences''.
\item \emph{Selective coding} to infer a central core category.
\item \emph{Validation} to confirm the developed theory with the participants.
\end{compactenum}

However, we already have a predefined set of codes (given RE problems) for which we want to know how the participants see their implications. For this reason,  we have to deviate our procedure from the standard procedure and rely on a mix of bottom-up and top-down. That is, we start with selective coding and build the core category with two sub-categories, namely \emph{RE problems} with a set of codes each representing one RE problem and \emph{Implications}, which then groups the codes defined for the answers given by the participants. For the second category, we conduct open coding and axial coding for the answers until reaching a saturation for a hierarchy of (sub-)categories, codes, and relationships.

\paragraph{Correlation Analysis for RQ~5}

As we are looking for general relationships between answers to our questions, we calculate Kendall's $\tau$ as correlation coefficient for ordinal data, i.e., on a Likert scale and binary answers coded as 0 and 1. As we are only interested in stronger relationships, we filter for $\tau$s larger than 0.5 and smaller than -0.5. The intuition is that we wanted at least a medium strength of the relationship as discussed in~\cite{cohen88}. We inspect each of the remaining relationships and test them for statistical significance with an $\alpha$ level of 0.05. Then, we discuss in the team of researchers if there is a reasonable interpretation for each of the remaining statistically significant correlation. If that is the case, we select it as candidate for inclusion in the theory.

\subsection{Validity Procedures}
\label{sec:ValidityProcedures}

As a means to increase the validity of the family of surveys, we have built the instrument on the basis of a theory induced from available studies (see Sect.~\ref{sec:Theory}). Furthermore, we conducted a self-contained, iterative validation phase before initiating the first survey in Germany (see Sect.~\ref{sec:Validation}). In particular, we conducted internal reviews and external reviews to increase the internal and the construct validity via researcher triangulation. To support for the external validity in advance, we conducted a pilot phase in an industrial context and used the feedback in further external reviews and dry-runs of the surveys. The external validity, however, will eventually be supported during replications that finally support empirical generalisations. 

However, as we plan to add the results from the correlation analysis and the Grounded Theory approach to our basic theory for future replications, we need to ensure the validity of especially those results. As shown in the previous section, we refer to researcher triangulation during the correlation analysis to decide for inclusion of the results. Regarding the results of applying Grounded Theory, we need to tackle the problem of expecting a small data population and not being able to validate our results with the participants, thus, leading to a possibly limited degree of saturation.

To tackle this problem, we rely also here in advance on researcher triangulation. To this end, we reproducibly perform the open coding and the axial coding in a team of two researchers. During open coding, we use spreadsheets to allocate the initial codes to the answers and to categorise the codes and their occurrences (e.g., ``change requests'' being named 4 times). We perform axial coding by following the relations from the single problems to the codes and counting also here the occurrences (e.g., ``change requests'' are named 3 times as a consequence of the problem ``hidden requirements'' and one time as a consequence of the problem ``moving targets''). After completing the open coding and the axial coding, a third researcher independently rebuilds a sample of codes and relationships on the basis of given statements to validate the results from the first run in the spreadsheet. This validation focusses on the occurrences of the codes rather than on the choice of the labels for the codes (e.g., ``CRs'' and ``change requests'' are seen as the same code). This shall allow to compensate for the potentially limited degree of saturation.

\section{Results from Germany}
\label{sec:Results}
In the following, we present the results from the first survey round conducted in Germany. We invited in total 105 contacts to participate in the survey as representatives for their companies. In cases of large enterprises with different business units focusing each on different application domains, we invited for each business unit one representative (if known). The contacts arise from previous research cooperations or knowledge transfer workshops for practitioners hosted at our universities.

In the following, we first summarise the information about the study population, before describing the results for each of the research questions. Questions for which we have no sufficient data are omitted (mostly additional open answer possibilities in MC questions). For the research questions 1--4, we first describe the results including descriptive statistics, before discussing the results in relation to our theory of Sect.~\ref{sec:Theory} (including the hypotheses tests). The analysis of the results to the open question in research question 4 (Grounded Theory) and research question 5 (correlation analysis) are summarised without interpretation in relation to our theory. 

\subsection{Study Population}

We registered 73 participants (out of the 105 invited ones) who did not, however, complete the full questionnaire. To get a consistent result set, we took 58 completed questionnaires into our result set. This gives us a response rate of 55~\%. The average time to complete the questionnaire was 28 minutes. This is similar to the expectation we had from pilot runs with about 30 minutes.

Most respondents (mode) work in an enterprise with more than 2,000 employees. The median are enterprises with
251--500 employees (Q~1). Therefore, the respondents tend to work in larger companies, but we have representatives from companies of all sizes. The respondents represent a broad range of software 
domains (see Tab.~\ref{tab:study-population}).

\begin{table}[htp]
\caption{Study population's software domains (Q~5).}
\centering \scriptsize
\begin{tabular}{p{0.55\linewidth}r}
\hline
Custom software development & 36 \% \\
IT consulting & 36 \%\\
Project management consulting & 35 \% \\
Software process consulting & 31 \% \\
Standard software development & 28 \%\\
Embedded software development & 7 \%\\
\hline
\end{tabular}
\label{tab:study-population}
\end{table}%
Most of the respondents (97~\%) work in companies that participate in globally distributed projects (Q~3). The large majority of respondents are located in Germany with a few exceptions located in Switzerland, Austria or France (Q~4). Most (mode) of the respondents work in the role of \emph{business analyst/requirements engineer} in their projects (Q~6). 80~\% of the respondents are experts with more than three years of experience. The rest has 1--3 years of experience (Q~7). The companies of the the respondents cover all the roles (customer, contractor, product development) in their projects. 19~\% state that they take the customer role, 47~\% take the role of a contractor, and 38~\% refer to product development (Q~8).

\subsection{Expectations on a Good RE (RQ~1)}
Regarding the practitioners' expectations on a good RE, we cover two topics: RE (process) improvement and expectations on RE company standards.

\subsubsection{RE Improvement} We first looked at how improving the RE compares to general software process improvements in other areas. Tab.~\ref{tab:beneficial-improvement} shows that the respondents considered process improvement in all offered areas as beneficial. We sort the results in this and the following tables with descending mode and then descending median.
\begin{table}[htb]
\caption{How beneficial would you personally rate an
improvement \ldots in your company? (Q~9: Not beneficial
at all: 1 \ldots Very beneficial: 5)}
\begin{center} \scriptsize
\begin{tabular}{p{0.35\linewidth}ccccc}
\hline
\textbf{Phase/discipline} & \textbf{Mode} & \textbf{Med.} & \textbf{MAD} & \textbf{Min.} & \textbf{Max.}\\
\hline
Requirements engineering & 5 & 5 & 1 & 2 & 5\\
Project management & 4 & 4 & 1 & 2 & 5\\
Architecture and design & 4 & 4 & 1 & 2 & 5\\
Quality assurance & 4 & 4 & 1 & 1 & 5\\
Implementation & 3 & 4 & 1  & 2 & 5 \\
\hline
\end{tabular}
\end{center}
\label{tab:beneficial-improvement}
\end{table}%

Despite seeing benefit in the improvement of all phases, it was considered very beneficial only for RE. For all the results, the 
deviation was small. The results about how challenging an improvement is differ only slightly. Only for RE improvements, most respondents rated it as very challenging (Tab.~\ref{tab:challenge-improvement}). 
\begin{table}[hbt]
\caption{How challenging would you personally rate an
improvement of \ldots in your company? (Q~10: Not challenging
at all: 1 \ldots Very challenging: 5)}
\begin{center} \scriptsize
\begin{tabular}{p{0.35\linewidth}ccccc}
\hline
\textbf{Phase/discipline} & \textbf{Mode} & \textbf{Med.} & \textbf{MAD}  & \textbf{Min.} & \textbf{Max.}\\
\hline
Requirements engineering & 5 & 4 & 1  & 1 & 5\\
Quality assurance & 4 & 4 & 1 & 1 & 5\\
Architecture and design & 4 & 3&1 & 1 & 5 \\
Project management & 4 & 3 & 1 & 1 & 5\\
Implementation & 4 & 3 & 1 & 1 & 5  \\
\hline
\end{tabular}
\end{center}
\label{tab:challenge-improvement}
\end{table}%

\begin{figure}[hbt]
\centering
  \includegraphics[width=.6\textwidth]{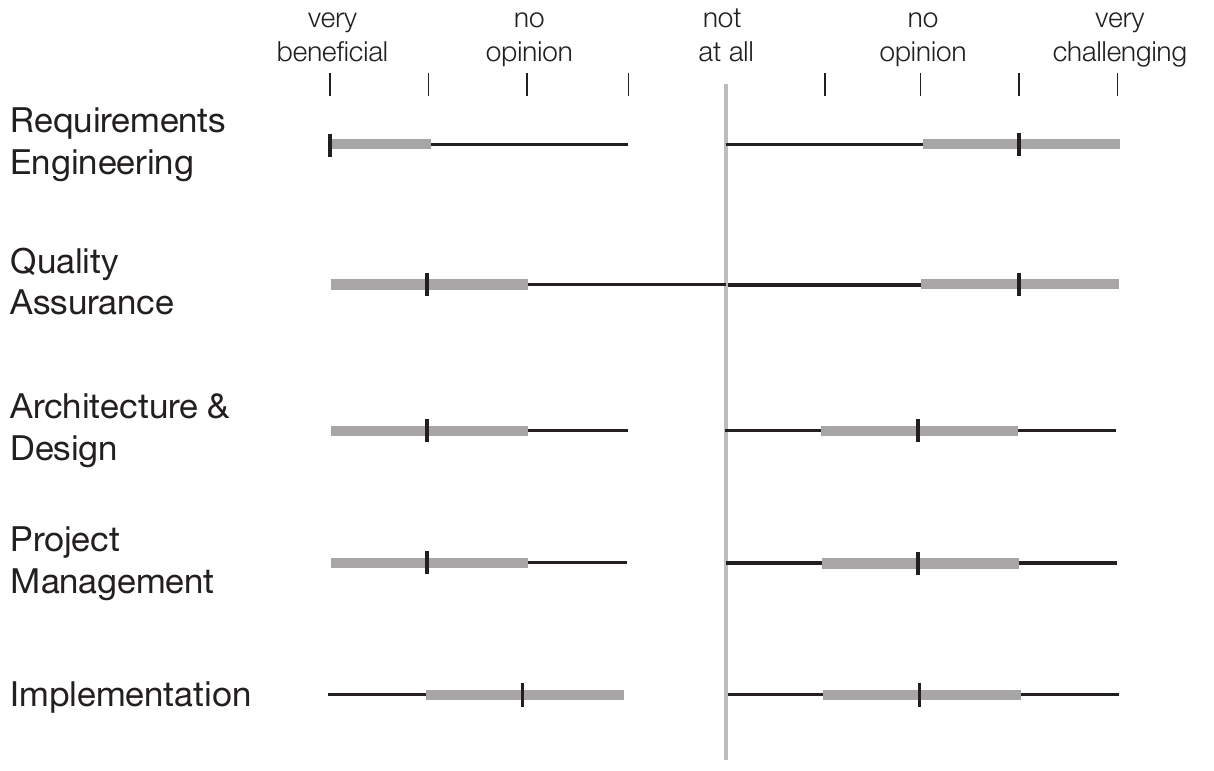}\\
  \caption{Comparision of beneficial (left side) and challenging (right side) areas for improvement. The vertical ticks show the median, the grey bars the MAD
  around the median and the horizontal lines the total range of answers.}\label{fig.bichart-beneficial-challenging}
\end{figure}

Again, however, all disciplines were considered as more or less challenging. The deviation here was also small. Fig.~\ref{fig.bichart-beneficial-challenging} shows a comparison of both the ratings for how beneficial and how challenging improvements in the different disciplines are. 

\subsubsection{Requirements Engineering Standard} 

\begin{table}[!htb]
\caption{Please rate the following statements on RE according to your expectations. (Q~11: I disagree: 1 \ldots I agree: 5)}
\begin{center}\scriptsize
\begin{tabular}{p{0.6\linewidth}ccccc}
\hline
\textbf{Statement} & \textbf{Mode} & \textbf{Med.} & \textbf{MAD}  & \textbf{Min.} & \textbf{Max.} \\
\hline
The standardisation of Requirements Engineering improves the overall process quality & 5 & 4 & 1 & 1 & 5\\
Offering standardised document templates and tool support benefits the communication  & 5 & 4 & 1 & 1 & 5 \\
Offering standardised document templates increases the quality of the work products & 4 & 4 & 1 & 1 & 5\\
The structure of documents should be standardised across different project environments, but the process itself should be left open for project participants & 4 & 4 & 1 & 1 & 5\\
The standardisation of Requirements Engineering hampers the creativity & 1 & 2 &1  & 1 & 5\\
\hline
\end{tabular}
\end{center}
\label{tab:general-statements}
\end{table}%

We then asked about the expectations of the respondents on standards in RE. The results are shown in Tab.~\ref{tab:general-statements}. On average, we see an agreement or moderate agreement on most statements we offered: The standardisation of RE improves the overall process quality. Offering standardised
document templates and tool support seems to benefit the communication and increases the
quality of the artefacts. The structure of documents should be standardised across
different project environments, but the process itself should be left open for project
participants. The only statement that received on average moderate disagreement was
that the standardisation of RE hampers the creativity. For all statements, the deviation
in the answers was low (MAD: 1). 

\begin{table}[!htb]
\caption{How important would you consider \ldots when defining an 
RE standard? (Q~12: Not important at all: 1 \ldots Very important: 5)}
\begin{center} \scriptsize
\begin{tabular}{p{0.35\linewidth}ccccc}
\hline
\textbf{Aspect} & \textbf{Mode} & \textbf{Med.} & \textbf{MAD}   & \textbf{Min.} & \textbf{Max.} \\
\hline
Support for agility & 5 & 4 & 1 & 1 & 5 \\
Definition of tailoring mechanisms & 5 & 4 &1 & 1 & 5\\
Definition of artefacts & 4 & 4 & 1  & 2 & 5\\
Definition of roles & 4 & 4 & 1 & 1 & 5\\
Definition of methods & 4 & 4 & 1 & 2 & 5\\
Support of impact analysis & 4 & 4 & 1 & 2 & 5\\
Process integration & 4 & 4 & 1 & 1 & 5\\
Support for prototyping & 4 & 4 & 1 & 1 & 5\\
Tool support for V\&V & 4 & 3.5 & 0.5  & 1 & 5 \\
\hline
\end{tabular}
\end{center}
\label{tab:important-aspects}
\end{table}%
Building on that, we asked about how important different aspects of a potential company-specific standard are. The results in Tab.~\ref{tab:important-aspects}
show that all the offered aspects seem to be important. 
The support for agility and the definition of a tailorable standard received the highest number of \emph{Very important} answers. All other aspects, i.e.\ the definition of artefacts, roles, methods, tool support for V\&V and the integration with other phases, the most common answers were \emph{Important}. The deviation was again low with a MAD between 0.5 and 1.

When asked about the motivation for a company-wide standard for
RE, the respondents agreed moderately with most of the given reasons
as shown in Tab.~\ref{tab:motivation-standard}. Exceptions are \emph{Better
quality assurance of artefacts} that received mostly agreements and
\emph{Formal prerequisite in my domain} that mostly received disagreement.
The deviation in all reasons was 1 or lower. 

\begin{table}[!htb]
\caption{Which reasons do you agree with as a motivation to define an RE standard?
(Q~13: I disagree: 1 \ldots I agree: 5)}
\begin{center} \scriptsize
\begin{tabular}{p{0.35\linewidth}ccccc}
\hline
\textbf{Aspect} & \textbf{Mode} & \textbf{Med.} & \textbf{MAD}  & \textbf{Min.} & \textbf{Max.}\\
\hline
Better QA of artefacts & 5 & 4 & 1  & 1 & 5 \\
Seamless development & 4 & 4 & 1  & 1 & 5\\
Better tool support & 4 & 4 & 1  & 1 & 5\\
Better progress control & 4 & 4 & 1  & 1 & 5\\
Higher efficiency & 4 & 4 & 1   & 1 & 5 \\
Knowledge transfer & 4 & 4 & 1  & 1 & 5\\
Support of project mgmt. & 4 & 4 & 0  & 1 & 5\\
Support of distributed dev. & 4 & 3 & 1  & 1 & 5\\
Support for benchmarks & 4 & 3 & 1  & 1 & 5\\
Compliance to regulations & 3 & 3 & 1   & 1 & 5 \\
Prerequisite in domain & 1 & 2 & 1   & 1 & 5  \\
\hline
\end{tabular}
\end{center}
\label{tab:motivation-standard}
\end{table}%

When asked about barriers to defining a company-wide standard for RE, the respondents mostly were neutral to
our proposed reasons (Tab.~\ref{tab:barrier-standard}). 
Only the missing willingness for change in the company was agreed by most
of the respondents, whereas the respondents disagreed with the lower efficiency.

\begin{table}[htp]
\caption{Which reasons do you see as barrier to define an RE standard?
(Q~14: I disagree: 1 \ldots I agree: 5)}
\begin{center} \scriptsize
\begin{tabular}{p{0.35\linewidth}ccccc}
\hline
\textbf{Aspect} & \textbf{Mode} & \textbf{Med.} & \textbf{MAD}  & \textbf{Min.} & \textbf{Max.}\\
\hline
Missing willingness for change & 5 & 4 & 1& 1 & 5  \\
Higher process complexity & 3 & 3 & 1 &1 & 5 \\
Higher communication demand & 3 & 3 & 1&1 & 5 \\
Missing possibility for standardisation & 3 & 3 & 1& 1 & 5\\
Lower efficiency & 1 & 2 & 1 & 1 & 5 \\
\hline
\end{tabular}
\end{center}
\label{tab:barrier-standard}
\end{table}%

\subsubsection*{Relation to our Theory} 
We repeat the hypotheses of our theory together with the p-values of the Wilcoxon tests in 
Tab.~\ref{tab:results-hypothesisRQ1}. We could fully support our theory that all phases or
disciplines are beneficial to improve. This is not the case for how challenging such an improvement
is. Only the hypotheses that RE and quality assurance improvement is challenging could be
supported. Hence, in that respect, RE seems to be especially challenging. Overall, we found
the results are too broad to add a lot of new information and, hence, will not test the hypotheses
in future replications.

\begin{table}[!htb]
\centering \scriptsize
\caption{Results of hypothesis tests for RQ~1. Statistically significant hypotheses highlighted (*).}
\label{tab:results-hypothesisRQ1}
\begin{tabular}{llp{0.7\linewidth}r}
\hline
\textbf{No.}& & \textbf{Hypotheses on improvement (Q~9 and Q~10)} & \textbf{p-value} \\ \hline 
H~1       & * & The improvement of all development phases is beneficial. & \\ 
H~1-a    &*  &  The improvement of RE is beneficial. & $<0.0001$\\ 
H~1-b    &*  &  The improvement of project management is beneficial. & $<0.0001$\\ 
H~1-c    &*  &  The improvement of architecture and design is beneficial. & $<0.0001$\\ 
H~1-d    &*  &  The improvement of implementation is beneficial. & 0.0006\\ 
H~1-e    &*  & The improvement of quality assurance is beneficial. & $<0.0001$\\ 
H~2   &   &  The improvement of all development phases is challenging. &  \\ 
H~2-a     &*  &  The improvement of RE is challenging. & 0.0002 \\
H~2-b    &  &  The improvement of project management is challenging. & 0.0898 \\ 
H~2-c     & &  The improvement of architecture and design is challenging. & 0.0580 \\ 
H~2-d     & &  The improvement of implementation is challenging. & 0.9067 \\ 
H~2-e   &*   &  The improvement of quality assurance is challenging. & 0.0060 \\ \hline
 &  & \textbf{Hypotheses on general expectations on standardisation in RE (Q~11)} & \\\hline
H~3     &* &  The standardisation of RE does not hamper creativity. & 0.0001 \\ 
H~4   &   *& The standardisation of RE improves the overall process quality. & $<0.0001$ \\ 
H~5      &  *& Standardised document templates and tool support benefits the communication. & $<0.0001$ \\ 
H~6      &  *& The structure of documents should be standardised across different project environments, but the process itself should be left open for project participants. & 0.0017 \\ \hline
&  & \textbf{Hypotheses on particular aspects in company-specific RE standards (Q~12)} & \\\hline
H~7  &    &  The definition of standard methods and modelling techniques is not important. & 0.9999 \\
H~8    &  &  Tool support for V\&V of req. spec. is not important. & 0.9958 \\ 
H~9     &* & The definition of standardised RE artefacts with document templates and/or tool support is important. & $<0.0001$ \\
H~10      &*& Tailoring mechanisms according to project characteristics are important. & $<0.0001$ \\ 
H~11      &  *& The definition of roles and responsibilities is important. & 0.0001 \\ 
H~12      & *& Support of impact analysis is important. & $<0.0001$ \\ 
H~13      &  *& Deep integration with other disciplines is important. & 0.0005 \\ 
 &     &  \emph{Support of agility in the process is\ldots} & \\
H~14	&    &  \ldots important (in case of business information systems  -- Q~5). & 0.3804 \\ 
H~15	&*    & \ldots not important (in case of embedded systems -- Q~5). & 0.0094 \\
 &     &  Support of prototyping is\ldots & \\
H~16	&    & \ldots important (in case of embedded systems -- Q~5). & 0.1015 \\ 
H~17	&    & \ldots not important (in case of business information systems -- Q~5). & 0.5857 \\ \hline
& & \textbf{Hypotheses on reasons for defining a company-specific RE standard (Q~13)} & \\\hline
H~18   &*   &  Formal prerequisite for project acquisition does not motivate a standard. & 0.0026 \\
H~19   &    &  Support of progress control does not motivate a standard. & 0.9750 \\
H~20    &*  &  Seamless development by integrating RE into development process motivates a standard. & $<0.0001$ \\
H~21     &* &  Better tool support motivates a standard. & 0.0021 \\
H~22    &  &  Support of distributed development motivates a standard. & 0.1224 \\
H~23    &*  &  Better quality assurance of artefacts motivates a standard. & $<0.0001$ \\
H~24   &   &  Support of benchmarks motivates a standard. & 0.5950 \\
H~25    &*  &  Support of project management and planning motivates a standard. & $<0.0001$ \\
H~26      &*&  Higher efficiency motivates a standard. & $<0.0001$ \\
H~27      &  *& Knowledge transfer motivates a standard. & $<0.0001$ \\ 
&    &  Compliance to regulations and standards (like CMMI)\ldots & \\
H~28	 &    &  \ldots does not motivate a standard in case of small/medium-sized companies  -- Q~1. & 0.4250 \\ 
H~29	&    &  \ldots motivates a standard in case of large companies (more than 2000 employees)  -- Q~1. & 0.1569 \\\hline
& & \textbf{Hypotheses on barriers for defining a company-specific RE standard (Q~14)} & \\\hline
H~30   &   &  Higher demand for communication does not barrier defining a standard.  & 0.3132\\
H~31   &*   &  Higher process complexity barriers defining a standard.  & 0.0416\\
H~32  &    &  Lower efficiency barriers defining a standard. & 0.9980\\
H~33    &*  &  Missing willingness to change barriers defining a standard. & 0.0041\\
H~34   &   &  Missing possibilities of standardisation barrier defining a standard. & 0.5662\\ \hline
\end{tabular}
\end{table}

We could also fully support our theory on general expectations on standardisation of RE (H~3--H~6). With strong statistical significance,
we corroborate that the standardisation of RE improves the overall process quality but does not hamper
the creativity. Standardised document templates and tool support benefits the communication. The
structure of the documents should be standardised across different project environments, but the
process itself should be left open for project participants. We will leave these hypotheses in the theory
for future testing.

Next, we tested our theory on specific aspects in company-specific RE standard (H~7--H~17). This led to
mixed results. We corroborated the hypotheses that the definition of standardised RE artefacts with
document templates and/or tool support is important, that tailoring mechanisms according to project 
characteristics are important, and that the definition of roles and responsibilities is important. Furthermore,
we found statistically significant that the support of impact analysis is important and that deep integration with
other disciplines is important. We split the hypothesis about the support of agility into two hypotheses depending
on the application area. We could only support that for embedded systems, the support of agility is not important.
We consider the rejection of H~14 as a particularity of German companies and, hence, leave this hypothesis in
the theory. For the hypotheses H~16 and H~17, we believe that our differentiation along the software domains
might not be appropriate. We will not split the hypothesis in the future theory and only assume that the support
of prototyping is important in general.

The hypotheses for H~18--H~29 are about reasons for defining a company-specific RE standard. The tests of
these hypotheses also show a mixed picture. We could corroborate the seamless development by integrating RE
into the development process, better tool support, better quality assurance of artefacts, support of project
management and planning, and higher efficiency and knowledge transfer to be motivational when establishing 
a standard. It is not a motivation for a standard that it is a formal prerequisite for project acquisition. Yet, we 
did not see a significant result about large companies having standards like CMMI as a motivation for an RE 
standard and smaller companies not. We tend to remove the split of these hypotheses and test in future
theories that compliance to regulations does not motivate an RE standard. For the support of progress control
 and benchmarks, we will adapt the theory according to the evidence into the opposite:
Support of progress control does motivate a standard. Support of benchmarks does not motivate a standard.
We observed several times, however that RE standards were motivated by supporting distributed development.
Hence, we decide to leave the corresponding hypothesis as is in the theory.

The last set of hypotheses for RQ~1 was about barriers for defining a company-specific RE standard. We could
only corroborate two hypotheses: Higher process complexity and missing willingness to change barrier defining
RE standards. As the rejected hypotheses were not clear from our experiences and literature, we decide to change
them according to the evidence into their opposites: Higher demand for communication barriers defining a
standard. Lower efficiency barriers defining a standard. Missing possibilities of standardisation barrier defining a standard.

\subsection{Status Quo in RE (RQ~2)}

\begin{figure}[hbt]
\centering
  \includegraphics[width=.65\textwidth]{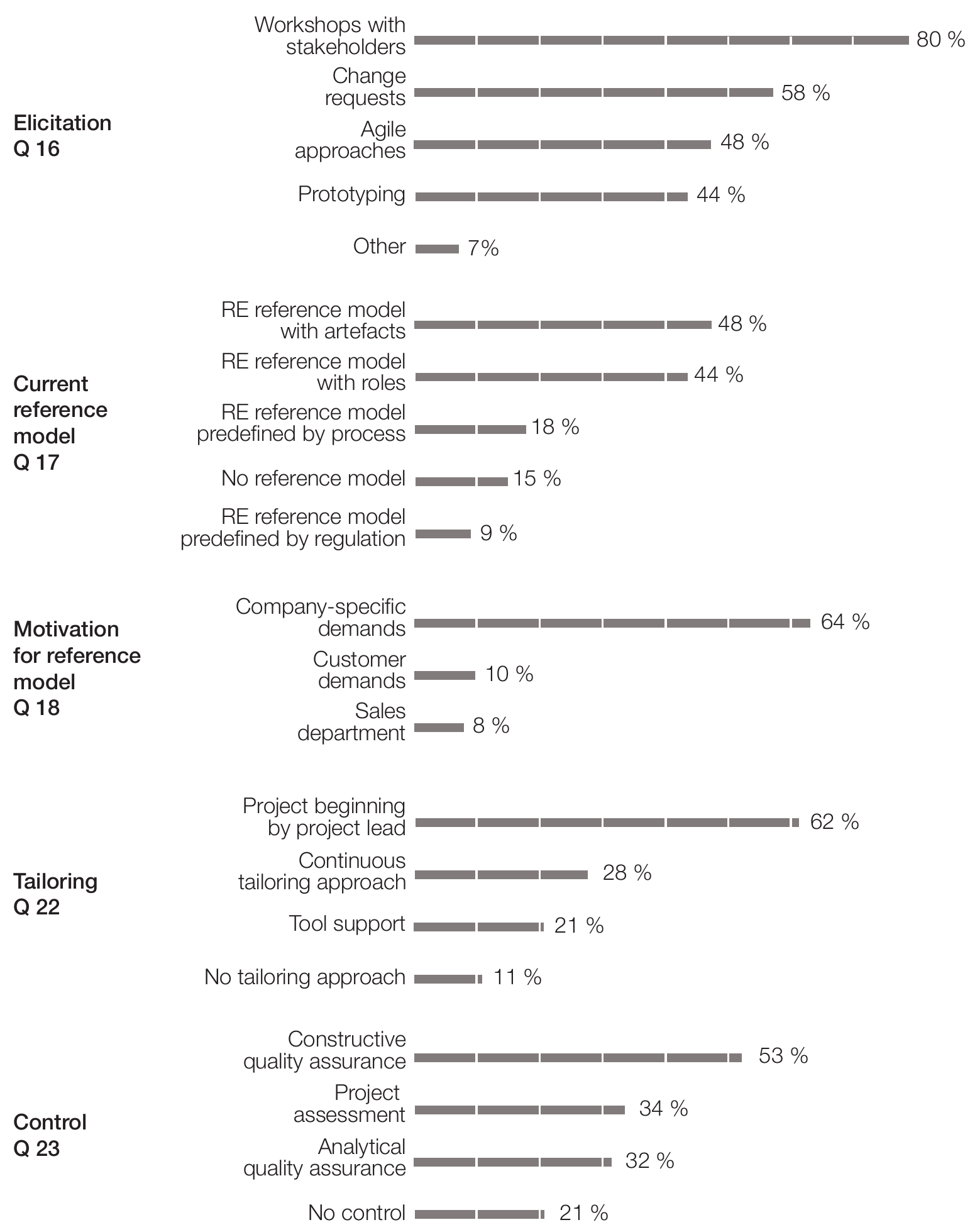}\\
  \caption{Summary of the status quo in RE. The bars show the percentage of the respondents who gave the corresponding answer.}\label{fig.boxplot-status-quo}
\end{figure}

After the expectations, we asked the respondents how they are involved
in RE in regular projects. An overview of the results is shown in Fig.~\ref{fig.boxplot-status-quo}.
Most of the respondents elicit and specify the
requirements themselves (Q~15). If they elicit requirements, we asked them about
how they elicit them. Of the respondents, 80~\% use workshops and discussions
with the stakeholders, 58~\% use change requests, 44~\% use prototyping, 48~\% refer to agile
approaches at the customer's site, and 7~\% use other approaches (Q~16).

Almost half of the respondents (44~\%) use an own RE standard that 
defines the process including roles and responsibilities. 45~\% of the respondents
put their focus on the definition of (coarse) artefacts and document templates. A standard that is 
predefined by the development process (e.g., Rational Unified Process) employ 18~\%, and 9~\% use a standard that is predefined according to a regulation (e.g., ITIL). Only 15~\% use no RE standard at all (Q~17).

The main reason for the definition of an RE standard were company-specific
demands (64~\%). Only 10~\% had an explicit demand from a customer and 8~\% because
of arguments from the sales department. Other reasons include \emph{to make requirements
more uniform} and \emph{quality and standardisation} (Q~18).

Overall, the respondents rate their RE standard well in terms of what it contains.
They mostly moderately agree with the statements about what their standard
contains (Tab.~\ref{tab:statements-standard}). Only the weaker statement that the model
has a differentiated view on different classes of requirements but not their dependencies
is rated mostly as an disagreement. By looking at the MAD, however, we observe that the deviation
for three of the statements is comparably high with 2. Hence, there is not a clear agreement between the
respondents.

\begin{table}[htp]
\caption{How would you rate \ldots to apply to your RE standard?
(Q 19: I disagree: 1 \ldots I agree: 5)}
\begin{center}\scriptsize
\begin{tabular}{p{0.4\linewidth}ccccc}
\hline
\textbf{Aspect} & \textbf{Mode} & \textbf{Med.} & \textbf{MAD}  & \textbf{Min.} & \textbf{Max.} \\
\hline
Classes of requirements \& dependencies & 4 & 3 & 1 & 1 & 5\\
Non-functional requirements & 4 & 3 & 2  & 1 & 5\\
Relies on architectural model & 4 & 3 & 2  & 1 & 5\\ 
Tracing & 4 & 3 & 2   & 1 & 5 \\
Classes of requirements, but no dependencies & 1 & 2 & 1 & 1 & 5\\ 
\hline
\end{tabular}
\end{center}
\label{tab:statements-standard}
\end{table}%

Change management of the requirements mostly falls into two categories: Almost half of the respondents (49~\%)
have a change management after the formal acceptance of the requirements specification and 45~\% have a continuous 
change management as part of an agile approach. 26~\% have change management during RE. Only 6~\% have
no change management at all (Q~20).

The majority of the respondents (53~\%) agreed that in their company each
project can decide whether to use the RE standard. That different business 
units have different standards as well as that all projects have to work according to the 
same standard each were agreed to by 30~\% (Q~21).

The tailoring of the RE standard is done at the companies of 62~\% of the respondents
at the beginning of the project by a project lead or a requirements engineer based
on experience. 28~\% have a tailoring approach that continuously guides the application
of the standard in their projects. 21~\% have tool support for tailoring their RE standard. 
11~\% state to not have a particular tailoring approach (Q~22).

We found similar rates for how the application of the RE standard is
controlled. 34~\% use project assessments, 32~\% use analytical quality assurance, e.g., as
part of quality gates, and 53~\% use constructive quality assurance, e.g., checklists or
templates. A fifth of the respondents (21~\%) do not control the application
of their RE standard at all (Q~23).

\subsubsection*{Relation to our Theory} 
We also tested our hypotheses for RQ~2 with the Wilcoxon test. The theories are shown in
Tab.~\ref{tab:results-hypothesisRQ2} with p-values. Our theories about the elicitation of requirements (H~35 and 
H~36) could both be supported. Requirements are elicited via workshops and change requests.  
We could also support both hypotheses on defining and tailoring RE standards. The standards are defined based on
demands specific for the company. The project lead tailors the standard at the beginning of a project based on experiences.

\begin{table}[!htb]
\centering \scriptsize
\caption{Results of hypothesis tests for RQ~2. Statistically significant hypotheses highlighted (*).}
\label{tab:results-hypothesisRQ2}
\begin{tabular}{llp{0.7\linewidth}r}
\hline
\textbf{No.} &  & \textbf{Hypotheses on the elicitation of requirements (Q~16)} & \textbf{p-value}\\\hline
H~35 & *     &  Requirements are elicited via workshops. & $<0.0001$ \\ 
H~36  & *    &  Requirements are elicited via change requests. & $<0.0001$ \\ \hline
& & \textbf{Hypotheses on defining and tailoring an RE standard (Q~18 and Q~22)} & \\\hline
H~37  &  *  &  Requirements engineering standards are defined due to company-specific demands. & $<0.0001$ \\ 
H~38  &  *    & \emph{The RE standard is tailored at the beginning of a project by the project lead based on experiences.} & $<0.0001$ \\ \hline
  & & \textbf{Hypotheses on the characteristics of their RE standard (Q~19)} & \\\hline
H~39   &   & The standard does not rely on an architecture model with different levels of abstraction & 0.6074 \\
H~40   &   &  The standard does not include a differentiated view on different requirements classes and dependencies. & 0.4625 \\ 
H~41 &     &  The standard does not include tracing relationships among the contents. & 0.2397 \\
H~42  &    &  The standard does not include a differentiated view on non-functional requirements. & 0.7982 \\  
H~43  &    & The standard includes a differentiated view on different requirements classes, but no dependencies. & 0.9969 \\ \hline
& & \textbf{Hypotheses on the application of the RE standard (Q~21 and Q~23)} & \\\hline
H~44  &  *    &  Each project can decide whether to use the standard. & $<0.0001$ \\ 
H~45   &  *   & The application of the RE standard is controlled via analytical quality assurance. & $<0.0001$ \\ \hline
& & \textbf{Hypothesis on change management (Q~20)} & \\\hline
H~46 &  *     &  A requirements change management is established after formally accepting a requirements specification. & $<0.0001$ \\ \hline
\end{tabular}
\end{table}

In contrast, our hypotheses about characteristics of company-specific RE standards were all not statistically significant. Hence,
the overall quality of the existing RE standard seems to be seen as higher than we expected. Hence, we decide to remove
H~43 as it is related to H~40 and adapt the other hypotheses (H~39--H42) to the opposite. 

We could, however, corroborate the hypotheses H~44 and H~45 about the application of RE standards. Each project can decide 
whether to use the company standard. Then, the application of the RE standard is controlled via analytical quality assurance.
Also the hypothesis about change management (H~46) was statistically significant: A requirements change management is established
after formally accepting a requirements specification. Hence, we will keep these hypotheses in the theory.

\subsection{Status Quo in RE Improvement (RQ~3)}

\begin{figure}[htbp]
\centering
  \includegraphics[width=.75\textwidth]{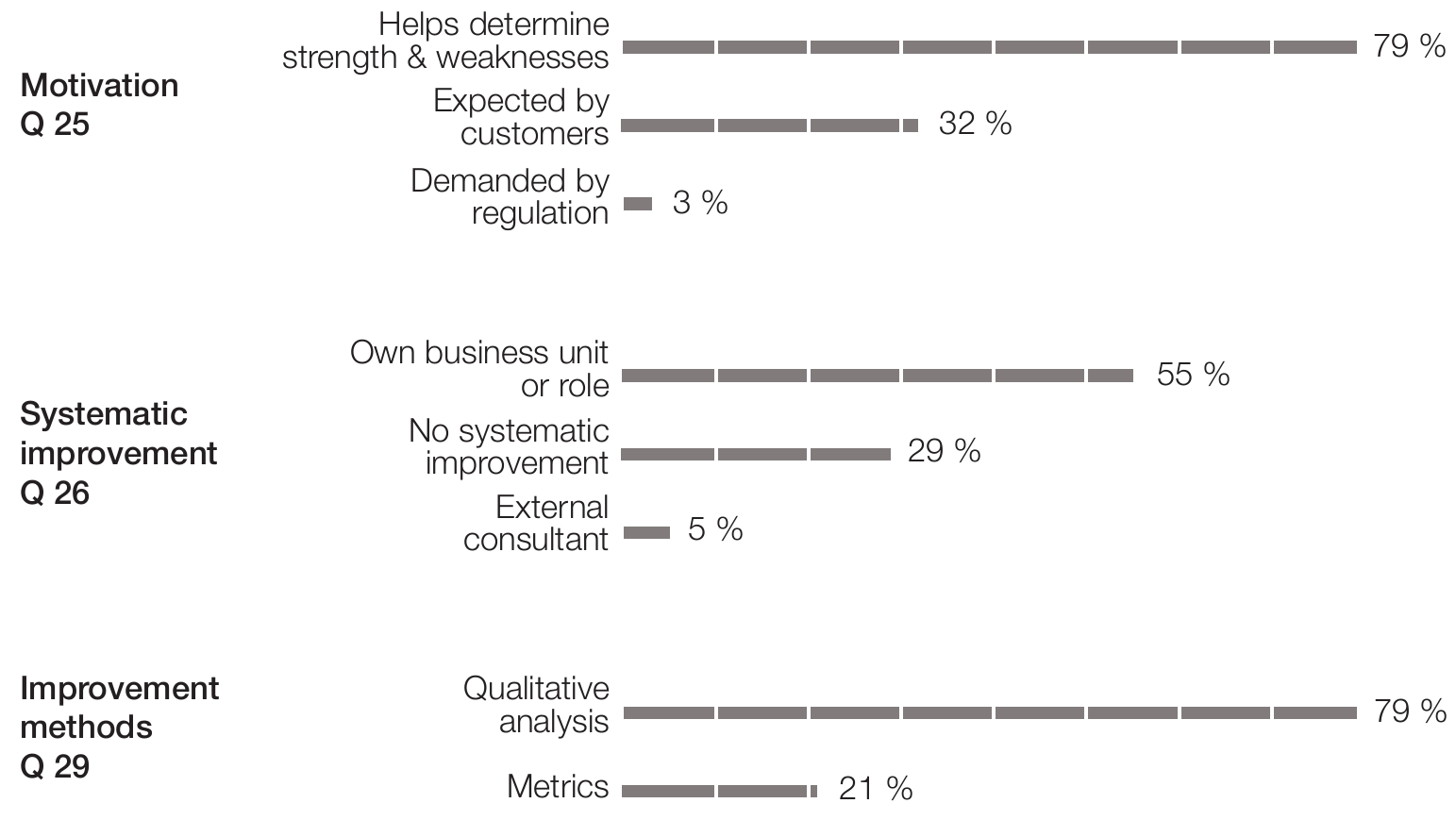}\\
  \caption{Summary of the status quo in RE improvement. The bars show the percentage of the respondents who gave the corresponding answer.}\label{fig.boxplot-status-pro-imp}
\end{figure}

Most of the respondents (83~\%) employ continuous improvement to RE (Q~24).  An overview of the
results on RE improvement is shown in Fig.~\ref{fig.boxplot-status-pro-imp}. When asked about
the motivation about this continuous improvement, of those, 79~\% think
that this continuous improvement helps them to determine their strengths and weaknesses and
to act accordingly. 32~\% agree that an improvement is expected by their customers. For only
3~\% of those with continuous improvement, it is demanded by a regulation (e.g., CMMI, Cobit
or ITIL) (Q~25). 

We then asked about how they conduct their RE improvements. 55~\% systematically improve RE
via an own business unit or role. 5~\% improve RE via an external consultant. 29~\% do not
systematically improve RE, but it remains the responsibility of the project participants.  Other
mentioned means to systematical improvements are an internal task force, retrospectives and
company-wide open space events (Q~26).

Half of the respondents with a continuous RE improvement (50~\%) explicitly state to not use a normative,
external standard for their improvement (Q~27). Several respondents use internal standards like an
internal process description system or best practices from literature. Exemplary statements for rejecting normative standards are ``I am not convinced of the external standards'', ``We want to live our own agility'', ``$[$We need more$]$ flexibility'', or ``$[$because of the$]$ individualism of the projects'' (Q~28).

Regarding the methods used, 79~\% of the respondents that employ improvement qualitatively analyse their projects, e.g., with interviews to gather lessons learnt.  21~\% refer to particular metrics and measurements to automatically assess their projects including customer satisfaction via A/B tests (Q~29, Q~30).

\subsubsection*{Relation to our Theory} 
Continuous improvement in RE seems to be performed in the majority of the companies. We found this also
to show in the result that our hypothesis that RE is not continuously improved was not significant 
(Tab.~\ref{tab:results-hypothesisRQ3}). Hence, we revert this hypothesis according to the empirical
evidence and will use \emph{Requirements engineering is continuously improved.} in the future theory.
We could, however, corroborate that if RE is continuously improved, it is done to determine strength and weaknesses. 
We will keep this hypothesis.

\begin{table}[!hbt]
\centering \scriptsize
\caption{Results of hypothesis tests for RQ~3. Statistically significant hypotheses highlighted (*).}
\label{tab:results-hypothesisRQ3}
\begin{tabular}{llp{0.7\linewidth}r}
\hline
\textbf{No.} & & \textbf{Hypotheses on continuous improvement (Q~24 and Q~25)} & \textbf{p-value} \\ \hline 
H~47    &   &  Requirements engineering is not continuously improved. & 0.4182 \\ 
H~48     & * &  A continuous improvement is done to determine strengths and weaknesses. & $<0.0001$ \\ \hline
&  & \textbf{Hypotheses on RE improvement (Q~26, Q~27, and Q~29)} & \\\hline
  &    & RE improvement is done systematically\ldots  & \\
H~49 &*	     &  \ldots via an own business unit/role in case of large companies (more than 2000 employees)  -- Q~1. & 0.0009\\ 
H~50	 &   &  \ldots via external consultants in case of large companies (more than 2000 employees)  -- Q~1. & 0.5000 \\ 
H~51   &*	    &  \ldots by project participants in case of small/medium-sized companies  -- Q1. & 0.0359 \\ 
H~52       &* & RE improvement is done via internally defined standards/approaches. & 0.0098 \\ 
H~53       & *& RE assessments/audits are done using qualitative methods. & $<0.0001$ \\ \hline

\end{tabular}
\end{table}

From our further hypotheses on RE improvement, we could support that RE is done systematically via an own business
unit / role in the case of large companies with more than 2,000 employees and by project participants in the case of small
and medium-sized companies. We could also corroborate that RE improvement is done via internally defined standards and
approaches and that RE assessments are done using qualitative methods. We will remove the hypothesis H~50 that 
RE improvement is done by external consultants in large companies as this seems to be not uniformly the case.

\subsection{Contemporary Problems in RE (RQ~4)}

Finally, after laying the groundwork about how RE is defined, lived and improved, we wanted to understand current problems in RE in practice. To this end, we first asked about the problems the participants experienced in their RE standards as well as the problems the participants experienced in their projects. The results are tested in relation to our theory introduced in Sect.~\ref{sec:Theory}. As a second step, we asked what implications the project-specific problems have and analysed the answers by applying Grounded Theory (see also Sect.~\ref{sec:DataAnalysis}). 

In the following, we introduce first the results in the stated problems and their relation to our theory, before introducing the results of what their implications are.

\subsubsection{Stated RE Problems}

Of our offered problems for the RE standards, most of them had a strong agreement among the respondents that they are not a problem (Tab.~\ref{tab:problems-standard}). Only two problems have as most frequent answer
\emph{I agree}: \emph{\ldots gives no guidance on how to create the specifications documents} and \emph{\ldots is not sufficiently integrated into risk management}. Both have lower medians, however, and the latter problem also has a high deviation. We also observed that for \emph{\ldots does not sufficiently define a clear terminology}, there is a higher median and also slightly increased deviation. Apart from these, the deviations are low for all problems.

\begin{table}[htb]
\caption{Please rate \ldots for your RE standard according to your experiences.
(Q~31: I disagree: 1 \ldots I agree: 5)}
\begin{center} \scriptsize
\begin{tabular}{p{0.5\linewidth}ccccc}
\hline
\textbf{Statement} & \textbf{Mode} & \textbf{Med.} & \textbf{MAD} & \textbf{Min.} & \textbf{Max.}  \\
\hline
Not integrated into risk management & 5 & 3 & 2 & 1 & 5\\
Gives no guidance on how to create specification documents & 4 & 2 & 1 & 1 & 5\\
Has no clear terminology & 3 & 3 & 2 & 1 & 5\\
Does not define roles & 2 & 2 & 1 & 1 & 5\\
Not integrated into design & 2 & 2  & 1 & 1 & 5 \\
Is too hard to understand & 1 & 2 & 1 & 1 & 4\\
Is too complex & 1 & 2 & 1 & 1 & 4 \\
Is too abstract & 1 & 2 & 1 & 1 & 4\\
Does not support precise specification & 1 & 2 & 1 & 1 & 5\\
Is too heavy weight & 1 & 2 & 1 & 1 & 4 \\
Is not flexible enough & 1 & 2 & 1 & 1 & 5\\
Does not allow for deviations & 1 & 2 & 1 & 1 & 5 \\
Not integrated into project management & 1 & 2 & 1 & 1 & 5\\
Not integrated into test management & 1 & 2 & 1 & 1 & 5\\
Does not scale  & 1 & 1.5 & 0.5 & 1 & 5\\
\hline
\end{tabular}
\end{center}
\label{tab:problems-standard}
\end{table}%

Second, we asked about more general problems in RE in the respondents' projects (Tab.~\ref{tab:problems-general}).
\begin{table}[htb]
\caption{How do \ldots apply to your projects?
(Q~32: I disagree: 1 \ldots I agree: 5)}
\begin{center} \scriptsize
\begin{tabular}{p{0.4\linewidth}ccccc}
\hline
\textbf{Problem} & \textbf{Mode} & \textbf{Med.} & \textbf{MAD} & \textbf{Min.} & \textbf{Max.}  \\
\hline
Separating requirements from solution & 5 & 4 & 1 & 1 & 5\\
Moving targets & 5 & 4 & 1 & 1 & 5 \\
Communication~flaws with the customer & 4 & 4 & 1 & 1 & 5\\
Incomplete and/or hidden requirements & 4 & 4 & 1 & 2 & 5\\
Inconsistent requirements & 4 & 4 & 1 & 1 & 5\\
Time boxing/not enough time & 4 & 4 & 1 & 1 & 5 \\
Underspecified requirements & 4 & 3.5 & 1.5 & 1 & 5\\
Communication~flaws within the team & 4 & 3 & 1 & 1 & 5\\
Terminological problems & 4 & 3 & 1 & 1 & 5\\
Gold plating & 4 & 3 & 1 & 1 & 5\\
Weak access to customer needs & 4 & 3  & 1 & 1 & 5\\
Unclear responsibilities & 3 & 3 & 1 & 1 & 5 \\
Insufficient support by project lead & 3 & 3 & 1 & 1 & 5\\
Insufficient support by customer & 3 & 3 & 1 & 1 & 5\\
Missing traceability & 3 & 3 & 1 & 1 & 5\\
High degree of innovation & 3 & 3 & 1 & 1 & 5 \\
Unclear non-functional requirements & 3 & 3 & 1 & 1 & 5 \\
Volatile customer business & 3 & 3 & 1 & 1 & 5\\
Technically unfeasible requirements & 2 & 2 & 1 & 1 & 5\\
Weak relationship to customer & 1 & 2.5 & 1.5 & 1 & 5\\
Weak knowledge of application domain & 1 & 2 & 1 & 1 & 5 \\
\hline
\end{tabular}
\end{center}
\label{tab:problems-general}
\end{table}%
There, we received a more mixed picture of RE in practice. On the one hand, the respondents agreed with problems like customers struggling with separating requirements from solutions and moving targets (see upper part of table) or moderately agreed with problems such as time boxing or gold plating. On the other hand, two problems were disagreed with: ``weak knowledge of the customer's application domain'', and ``weak relationship to customer'' (lower part of table). The rest was considered neutral or was moderately disagreed with. The deviations were mostly small (1). In two problems, we have a deviation of 1.5, which suggests a slightly higher diversity in the answers. Accordingly, we could not find any large or significant correlation between the problems stated by the respondents and the answers selected in RQ~1 and 2 (e.g., the relevance given to the artefacts and problems experienced w.r.t. inconsistency). 

\begin{figure}[!hptb]
\centering
  \includegraphics[angle=90, width=.7\textwidth]{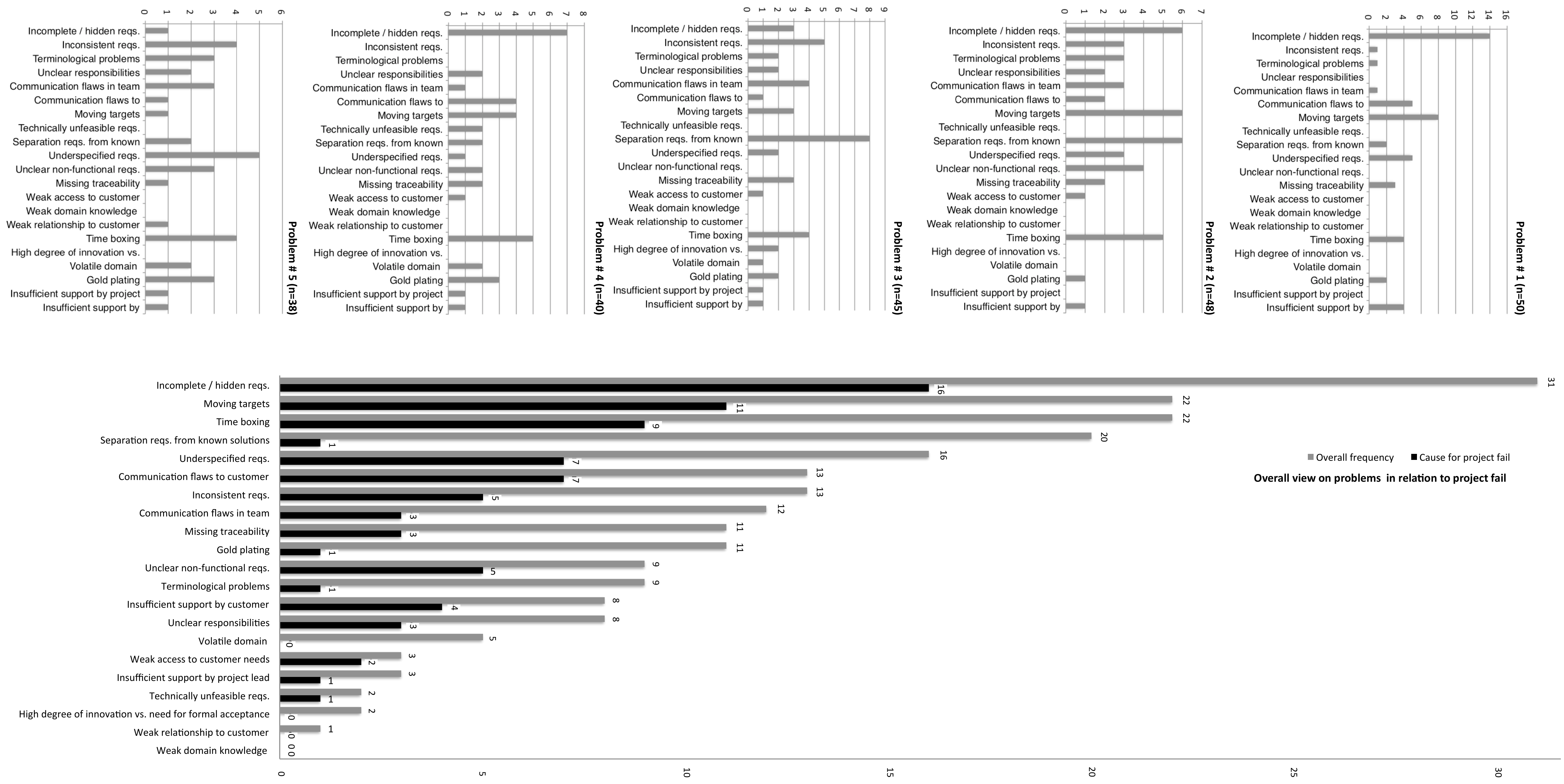}\\
  \caption{Summary of selected most critical problems and their causes for project failures (Q~33, Q~35).}
  \label{fig.problemssummary}
\end{figure}

Finally, we asked the respondents to rank the problems as they have experienced them according to their criticality from 1 to 5 (Q~33).  The results of the ranking are illustrated in Fig.~\ref{fig.problemssummary}. On the left side of the figure, we illustrate the ranking of the problems according to their criticality in the sense of their frequency (e.g., ``problem number 1''). The right side of the figure accumulates all problems according to their overall frequency and additionally shows how often the participants have stated those problems to be a reason for a project fail in their experience (Q~35). The most often mentioned of the most critical problems in this ranking are incomplete and/or hidden requirements and moving targets. Also mentioned often are time boxing/not enough time. Please note, however, that our results on project failures do not yet allow for a clear interpretation as not all respondents selected five of the given problems.

\subsubsection*{Relation to our Theory} 
As non of our offered problems with RE standards were mostly agreed with, we could only support few
of our hypotheses in this area (Tab.~\ref{tab:results-hypothesisRQ4}). We only could corroborate that the
RE standard is not too hard to understand and not too complex. It scales to the high complexity of the
respondents' projects and it sufficiently defines roles and responsibilities. We expected significant problems
in the other hypotheses which does not seem to be the case. As we have seen several of these problems
in practice, we decide to leave them in the theory for further replication but to augment them with additional
problems with RE standards to get a clearer picture if there are really no problems.

\begin{table}[!htb]
\centering \scriptsize
\caption{Results for hypothesis tests for RQ~4. Statistically significant hypotheses highlighted (*).}
\label{tab:results-hypothesisRQ4}
\begin{tabular}{llp{0.7\linewidth}r}
\hline
\textbf{No.} & & \textbf{Hypotheses on problems with the RE standard (Q~31)} & \textbf{p-value}\\\hline
H~54     &*  & It is not too hard to understand. & $<0.0001$ \\ 
H~55      &*  & It is not too complex. & 0.0098 \\ 
H~56      &* & It scales to our projects' high complexity. & $<0.0001$ \\ 
H~57     &*  & It sufficiently defines roles and responsibilities. & 0.0001 \\ 
H~58     &  & It is too abstract. & 0.9999 \\ 
H~59      & & It does not support the specification of precise requirements. & 0.9999 \\ 
H~60     &  & It is too heavy weight for our projects. & 0.9999 \\ 
H~61     &  & It is not flexible enough. & 0.9998 \\ 
H~62     &  & It does not sufficiently define a clear terminology. & 0.8201 \\ 
H~63     &  & It gives no guidance on how to create the specification documents. & 0.9979 \\ 
H~64     &  & It does not sufficiently allow for deviations according to project circumstances. & 0.9999 \\ 
H~65     &  & It isn't sufficiently integrated into project management. & 0.9932 \\ 
H~66     &  & It isn't sufficiently integrated into design and architecture. & 0.9868 \\ 
H~67     &  & It isn't sufficiently integrated into risk management. & 0.3548 \\ 
H~68     &  & It isn't sufficiently integrated into test management. & 0.8209 \\ \hline
& & \textbf{Hypotheses on RE problems within projects (Q~32)} & \\\hline
H~69    &  & No communication flaws within the project team are not a problem. & 0.5121 \\ 
H~70     &*  & Insufficient support by project lead is not a problem. & 0.0010 \\ 
H~71      & & Gold plating (implementing features without corresponding requirements) is not a problem. & 0.5985 \\ 
H~72    &*  & Weak knowledge of the customer's application domain is not a problem. & $<0.0001$ \\ 
H~73      &* & Technically unfeasible requirements is not a problem. & 0.0004 \\ 
H~74      &* & A weak relationship to the customer is not a problem. & $<0.0001$ \\ 
H~75      &*  & A volatile customer's business domain regard, e.g., changing business requirements, is not a problem. & 0.0204 \\ 
H~76     &* & Communication flaws between project team and customer are a problem. & 0.0007 \\ 
H~77     & & There are terminological problems. & 0.1784 \\ 
H~78     & & Unclear responsibilities are a problem. & 0.3143 \\ 
H~79     &* & Incomplete and/or hidden requirements are a problem. & $<0.0001$ \\ 
H~80      && Insufficient support by the customer is a problem. & 0.1696 \\ 
H~81    &*  & Stakeholders with difficulties in separating requirements from previous solutions are a problem. & 0.0002 \\ 
H~82     &*  & Inconsistent requirements are a problem. & $<0.0001$ \\ 
H~83    &   & Missing traceability is a problem. & 0.3165 \\ 
H~84    &*  & Moving targets are a problem. & 0.0003 \\ 
H~85    & & Weak access to customer needs and/or business information is a problem. & 0.2173 \\ 
H~86      &* & Time boxing / Not enough time in general is a problem. & 0.0154 \\ 
H~87     &  & The discrepancy between the high degree of innovation and the need of formal requirements acceptance is a problem. & 0.1224 \\ 
H~88      &* & Underspecified requirements that are too abstract and allow for various interpretations are a problem. & 0.0109 \\ 
H~89      &* & Unclear/unmeasurable non-functional requirements are a problem. & 0.0148 \\ \hline
\end{tabular}
\end{table}

We could, however, support our theory on general problems in RE in several cases. Most of the mentioned problems
are statistically significant for the respondents. We had to refute that communication flaws within the
project team and gold plating are no problems. We could also not support that terminological problems, unclear responsibilities, 
missing traceability, weak access to customer needs, and insufficient support by the customer were problems. Finally, it was not
significant that there is a discrepancy between high degree of innovation and the need of formal requirements acceptance.
Apart from that, we could corroborate several problems, such as moving targets, and others, e.g., that there is sufficient support by by the project lead. As all these hypotheses are based on our experiences and/or existing literature, we decide to keep all hypotheses on RE problems in project in the theory. We believe we need to test them in the next replications before we can decide to remove or change them.

\subsubsection{Implications of the Problems}

After asking for the appearance of given problems in the respondents' socio-economic contexts, we asked the implications those problems have. As discussed earlier, we analyse this question without any pre-defined expectations (theory) and applied Grounded Theory. Due to the resulting complexity in given answers and resulting coding scheme, we describe the results step-wise.

To this end, we first introduce the full results from the open coding, followed by the full results of the axial coding. In a last step, we present a condensed overall view of the results as a graph with a minimal saturation. That is, we show only those results having a minimal occurrence in the answers to include only those in our theory.

\subsubsection*{Full Results from Open Coding}
\begin{figure}[!htb]
\centering
  \includegraphics[width=0.85\textwidth]{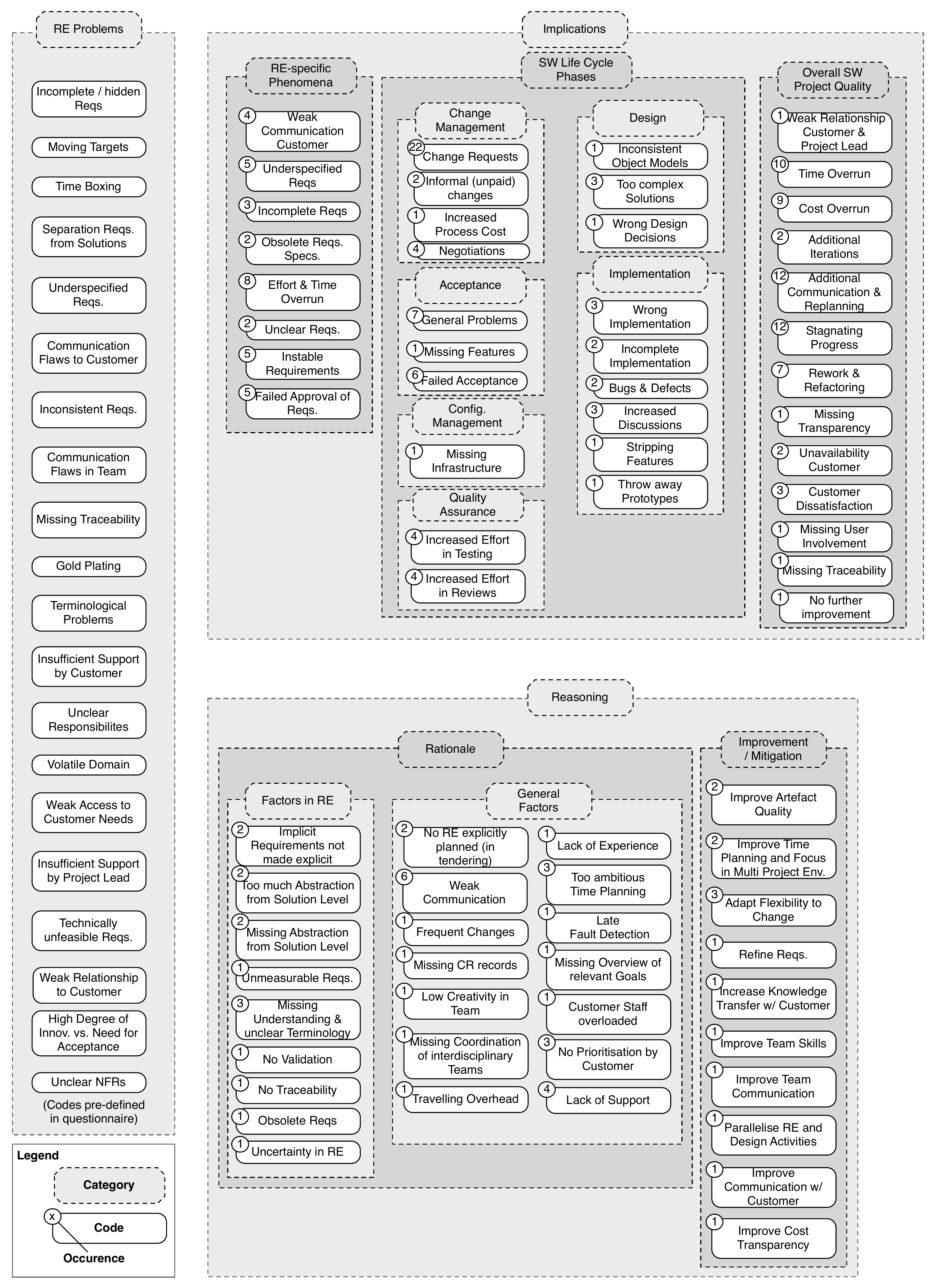}\\
  \caption{Categories and codes resulting from open coding.}
  \label{fig.OpenCoding}
\end{figure}

Figure~\ref{fig.OpenCoding}  summarises the results from the open coding. We distinguish a hierarchy of categories as an abstraction of those codes defined for the answers given in the questionnaire. For each code, we furthermore denote the number of occurrences. Not included in the coding are statements that cannot be unambiguously allocated to a code, for example, the statement ``never ending story'' as an implication of the problem ``Incomplete / Hidden requirements''.

Given that we asked what implications the RE problems (resulting from our theory) have, we would expect two top-level categories. The participants also stated, however, reasons for the problems and occasionally also how they would expect to mitigate the problem. As shown in Fig.~\ref{fig.OpenCoding}, we thus categorise the results into the pre-defined category \emph{RE Problems}, \emph{Implications}, and the additional category \emph{Reasoning}.

Regarding the implications, we distinguish three sub-categories. Consequences of problems in the RE phase itself, consequences to be seen in further phases of the SW life cycle other than RE, and more abstract consequences on the overall project quality. The highest occurrence of statements is given for the code \emph{Change Request} being stated 22 times. Other codes resulted from only one statement, but they were unique, specific formulations that could not be merged with other statements. For instance, the code \emph{Weak Relationship Customer \& Project Lead} in the category \emph{Overall SW Project Quality} resulted from a statement, which we could not allocate to another code without interpretation and potentially misinterpreting the statement (given that no validation with the respondents is possible).

Regarding the reasoning for given RE problems, we distinguish the category \emph{Rationale} as a justification of why particular problems occurred, and \emph{Improvement/Mitigation} for statements that suggested how to mitigate particular problems.  The first category can be further divided into \emph{Factors in RE} and \emph{General Factors}. 

Also here, we encountered very mixed statements including detailed ones we had to allocate to codes having in the end only one occurrence and vague statements we could accumulate with (consequently vague) codes. Subsequent original statements shall give an impression about the answers given:
\begin{compactitem}
\item[Code \emph{Missing Abstraction from Solution Level:}] ``Stakeholders like to discuss on solution level, not on requirements level. Developers think in solutions. The problem is: even Product Managers and Consultants do it.''
\item[Code \emph{No RE explicitly planned (in tendering):}] ``A common situation is to take part in a tender process -- where requirements are specified very abstract -- most of these tender processes do not include a refinement stage, as a supplier we are bound to fulfill vague requests from the initial documents.''
\item[Code \emph{Weak Communication:}] ``The communication to customer is done not by technicians, but by lawyers.''
\item[Code \emph{Too Ambitious Time Planning:}] ``Delivery date is known before requirements are clear.''
\item[Code \emph{Implicit Requirements not made explicit:}] ``Referencing  common sense  as a requirement basis.''
\item[Code \emph{Failed Acceptance:}] ``After acceptance testing failed, the hidden requirements came up and had to be fixed on an emergency level.''
\item[Code \emph{Missing Coordination of Interdisciplinary Teams:}] ``Missing coordination between different disciplines (electrical engineering, mechanical engineering, software etc.).''
\end{compactitem} 

\subsubsection*{Full Results from Axial Coding}

The axial coding defines the relationships between the codes. As a consequence of the categories introduced in the previous section, we distinguish two types of relationships:
\begin{compactenum}
\item The consequences of given RE problems to the category \emph{Implications}, and
\item The consequences of the codes in the category \emph{Reasoning} to the RE problems including rationales and improvement suggestions.
\end{compactenum}

The full results of the axial coding can be taken from Fig.~\ref{fig.axialCoding1} and Fig.~\ref{fig.axialCoding2} in~\ref{sec:axialappendix}, and a detailed categorisation of the shown codes can be taken from the previous section (Fig.~\ref{fig.OpenCoding}). We further refrain from interpreting any transitive relationships from reasonings to implications because of the multiple input-/output-relationships between the codes of the different categories; for instance, while ``Too ambitious Time Planning'' was stated as an exclusive reason for ``Time Boxing'' as an RE problem, the problem ``Incomplete / hidden Requirements'' has multiple reasons as well as multiple consequences.

\subsubsection*{Condensed Overall View with Minimal Saturation}

\begin{figure}[!htb]
\centering
  \includegraphics[width=1\textwidth]{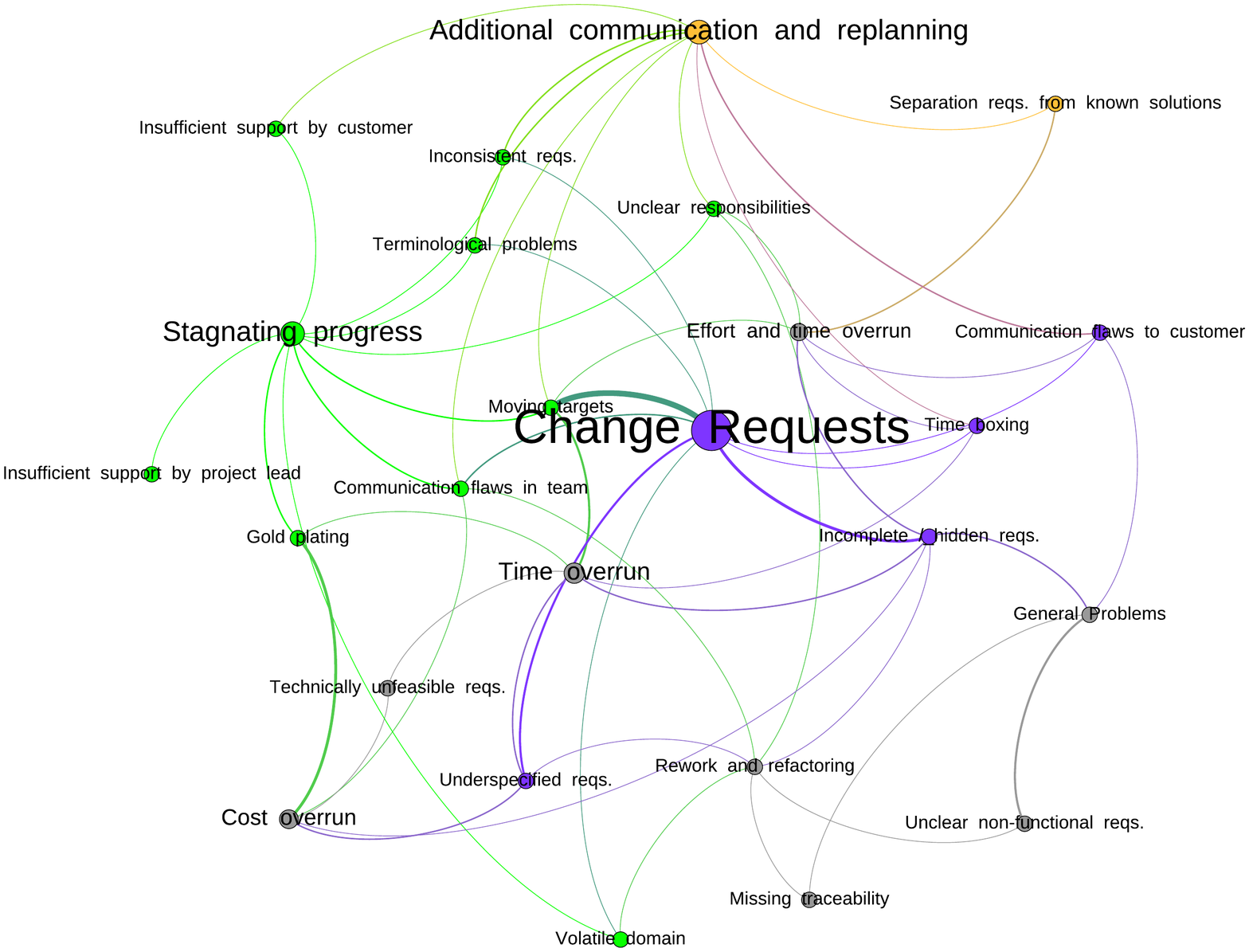}\\
  \caption{Condensed view on axial coding with minimal weighting of 7 in the nodes.}
  \label{fig.axialCodingCondensed}
\end{figure}

Finally, we are especially interested in a condensed result set that omits the codes and the dependencies with limited occurrences in corresponding statements. The reason is that we need a result set with a minimal saturation to propose its integration into our theory (Sect.~\ref{sec:Theory}) for the future replications of the survey.  After testing the results with different values for the minimal occurrences, we define a graph including only codes with a minimal occurrence level of 7. This resulting graph can be taken from Fig.~\ref{fig.axialCodingCondensed}. Nodes represent a selection of codes and edges represent a selection of relationships between nodes.

With the chosen minimal occurrence level, the final graph does not include statements coded in the category \emph{Reasoning} leaving us with a selection of RE problems interconnected with their implications. The three nodes with the highest occurrence in their underlying statements are differently coloured. Change requests (in the centre of the figure) are stated as the most frequent consequence of various RE problems such as time boxing or incomplete / hidden requirements. Additional communication and replanning (upper part of the figure) was another frequently stated consequence of interconnected RE problems, similar as a stagnating process (left side of the figure).

\subsection{Patterns of Expectations, Status Quo and Problems in RE (RQ~5)}

Finally, we checked if we can find interesting patterns of relationships in the answers we got for all questions. In the
following, we discuss all correlations between answers with an at least moderate, significant correlation for which
we found a reasonable interpretation. We excluded, for example, correlations between the questions about the
domain and application areas. We also excluded few correlations from other questions. For example, we could not
interpret that who has a change management approach that applies after formally accepting a requirements specification, 
uses qualitative methods for assessments. All remaining correlations are shown in Tab.~\ref{tab:results-correlation}.


\begin{table}[!htb]
\centering \scriptsize
\caption{Significant correlations with an at least moderate correlation coefficient.}
\label{tab:results-correlation}
\begin{tabular}{rp{0.7\linewidth}rr}
\hline
\textbf{No.}& \textbf{Correlation} & \textbf{Kendall} & \textbf{p-value} \\ \hline 
C~1 & Who does custom software development does not systematically improve. & 0.5529 & 0.0007\\
C~2 & Who rated improvement in architecture and design as beneficial also rated improvement in implementation as beneficial. & 0.5294 & $<0.0001$\\
C~3 & Who rated improvement in architecture and design as challenging also rated improvement in implementation as challenging. & 0.5301& $<0.0001$\\
C~4 & Who thinks that offering standardised document templates and tool support benefits the communication also thinks that offering standardised document templates increases the quality of the work products. & 0.5720 & $<0.0001$\\
C~5 & Who rates tool support for validation and verification of requirements specification as important also agrees with tool support as a motivation to define a company-wide standard. & 0.5215 & $<0.0001$\\
C~6 & Who rates the RE standard as too hard to understand also rates it as too complex. & 0.8286 & $<0.0001$\\
C~7 & Who rates the RE standard as too hard to understand also rates it as too abstract. & 0.7857 & $<0.0001$\\
C~8 & Who rates the RE standard as too complex, also rates it as too abstract. & 0.8082 & $<0.0001$\\
C~9 & Who rates the RE standard as too hard to understand, also rates it as not scaling to the projects' high complexity. & 0.5165 & $<0.0001$\\
C~10 & Who rates the RE standard as too complex, also rates it as not scaling to the projects' high complexity. & 0.5000 & $<0.0001$\\
C~11 & Who rates the RE standard as not scaling to the projects' high complexity, also rates it as too heavy-weight. & 0.5908 & $<0.0001$\\
C~12 & Who rates the RE standard as too heavy-weight, also rates it as not flexible enough. & 0.5001 & $<0.0001$\\
C~13 & Who rates the RE standard as not supporting the specification of precise requirements, rates it as not sufficiently defining terminology.
& 0.5228 & $<0.0001$\\
C~14 & Who rates the RE standard as not sufficiently defining a clear terminology, rates it also as giving no guidance on how to create specifications. & 0.5462 & $<0.0001$\\
C~15 & Who rates the RE standard as too hard to understand, also rates it not sufficiently allowing for deviations according to project circumstances that cannot be formalised. &
0.5092 & $<0.0001$\\
C~16 & Who rates the RE standard as too heavy-weight, also rates it as not sufficiently defining roles and responsibilities. &
0.5433 & $<0.0001$\\
C~17 & Who rates the RE standard as not sufficiently defining a clear terminology, also rates it as not sufficiently integrated with project management. & 0.5476 & $<0.0001$\\
C~18 & Wo rates the RE standard as not sufficiently integrated into project management, also rates it as not sufficiently integrated into design
and architecture. & 0.6292 & $<0.0001$\\
C~19 & Wo rates the RE standard as not sufficiently integrated into project management, also rates it as not sufficiently integrated into risk
management. & 0.6027 & $<0.0001$\\
C~20 & Who rates the RE standard as not sufficiently integrated into project management, also rates it as not sufficiently integrated into test
management. & 0.5765 & $<0.0001$\\
C~21 & Who rates the RE standard as not sufficiently integrated into risk management, also rates it as not sufficiently integrated into test
management. & 0.6542 & $<0.0001$\\
C~22 & Where time boxing / not enough time applies, also a discrepancy between high degree of innovation and need for formal acceptance applies. & 0.5253 & $<0.0001$\\
\hline
\end{tabular}
\end{table}

Correlation C~1 states that those who work in custom software development do not systematically improve
their RE processes. We explain this by the different environments and customer processes many in
custom development have to work in and, hence, a systematic improvement is often difficult. Correlations
C~2 and C~3 together suggest that those who rate architecture improvement as beneficial and challenging
also rate implementation improvement as beneficial and challenging. This is reasonable because architecture
and implementation are closely related and, therefore, improvements will be similar for both.

Correlation C~4 expresses that those who think that offering standardised document templates and tool support
benefits the communication also think that offering those templates increases the quality of the work products.
So, if standardised document templates are seen positive, both positive aspects we offered exist in practice.
Similarly, if tool support is seen as positive, it is also a reason to build an RE standard. This is expressed
in C~5 which says that those who rate tool support for V\&V of requirements specifications as important also
agree with tool support as a motivation to define a company-wide standard.

The following correlations C~6 to C~21 are all related to problems with the existing RE standard of the
respondents. We found many, partly very strong, correlations between these problems. Our abstract interoperation 
is that if a respondent has a problematic RE standard, it often has many problems. This is in contrast to the results
of the analysis of single questions where we found none of the problems to be significant. Hence, it seems either
their RE standard has a multitude of problems or none. Also if integration into other phases is a problem, it is
problematic for all phases (C~18--C~21). Several of these correlations seem directly logical. For
example, C~10 states that if an RE standard is too hard to understand, it is also too abstract. Moreover, we found 
the following two particularly interesting: C~10 suggest that those who rate the RE standard as too complex also
rates it as not scaling the complexity of the projects. We expected that complex standards at least are able to
better support complex projects. This seems not to be the case. C~14 states that those who rate the RE standard
as not sufficiently defining a clear terminology also rate it as giving no guidance on how to create specifications.
This fits well to our observations from practice that terminology is a core aspect of a standard if we want to
provide guidelines for creating artefacts.

Finally, C~22 says that where time boxing applies and there is not enough time in general, there is also a discrepancy
between the high degree of innovation and the need for formal acceptance. These two aspects fit well together: If
we need to formally accept the requirements in a context of high innovation and project management at the same time
gives only a pre-defined time box for RE, it will become very difficult to provide sensible requirements for acceptance.

\section{Lessons Learnt}
\label{LessonsLearnt}

In the course of planning the family of surveys and of executing the first survey presented in the article at hand, we collected a number of lessons learnt. In the following, we summarise most important lessons learnt and briefly describe those things we change to facilitate the ongoing and the future replications.

\paragraph{Transparency by Invitation} We intentionally opted for an invitation-based survey and had good results regarding the reproducibility of the response rate, but also regarding the amount of participants actually completing the survey despite the high amount of questions (with an average of 28 minutes needed to complete the survey). In consequence, however, many companies did not participate in the survey and even within the companies, we cannot guarantee that we always got a respondent representative for the company. Therefore, we still face a trade-off between the advantages and the disadvantages of conducting the survey by invitation only. For the future replications, we are continuing this modus while making the whole endeavour more public (e.g., with the website) where participants not considered in the invitations can request a code for the survey (which still allows us to track the number of participants). However, as we are consequently expecting more participants to which we have no relation at all, we need to make the questionnaire more concise and the collected data more reproducible (see also subsequent paragraphs).

\paragraph{Sample Representativeness} As we did not select the sample of invitees following a known distribution of context factors in Germany, we cannot guarantee for its representativeness. Although we carefully tracked our invitations, and thereby we are able to get a reasonably good response rate, we cannot explain, for example, the high number of consultants among the respondents. One explanation we have is that they are most interested in requirements engineering and that many companies do not explicitly define the role of a requirements engineer, but see RE-related tasks as consulting (potentially run as an own project). In general, the invitation list was more balanced between consultants and other software domains. In future replications, we need to take more care for ensuring the representativeness of the sample and are currently defining a set of context variables that allow to carefully check for the representativeness of our samples.

\paragraph{Conciseness of Questionnaire} One aspect we consider important for the replications is the conciseness of our questionnaire as, for the first round, it took an average of 28 minutes to fully answer the survey. At the same time, we need to maximise the construct validity, i.e. we need to gather as much information as possible to accurately answer our research questions. As already stated, we intentionally planned for each replication a planning step where we change the questionnaire, also according the results from the previous surveys. Currently, we are changing those questions that did not bring fruitful results or where we could not confirm corresponding hypotheses, we are identifying candidates for removal, and we are defining completely new questions. For example, we changed the initial questions regarding the industry sector following an international standard (see the previous paragraph), and we also changed the question on the requirements elicitation. Questions we already removed were, for example, to what extent projects in various business units have to follow a company standard as this questions was too specifically dedicated to large companies. An exemplary new question is how the respondents deal with changing requirements after the initial product release. 

\paragraph{Level of Detail in the Questionnaire} Despite the pilot runs and the current work on making the questionnaire more concise, we still might have questions that allow for misinterpretations; for example, different respondents might implicitly see the aspect \emph{traceability} from different angles (e.g., between requirements, from requirements to their rationale or from requirements to the code level). We are currently working on clarifying those potentially misleading variables by including examples into the questionnaire and making the answer possibilities more precise. Also, for the future replications, we need to collect more information about the status quo in requirements engineering, for example by asking what artefacts area created (e.g., use case models, goal models or a glossary) as the knowledge about the status quo allows for more investigations and reasoning about the problems the respondents might experience (e.g., via correlation analyse).

\section{Conclusion}
\label{sec:Conclusion}

In this article, we contributed the design of a global family of surveys to overcome the problem of by now isolated investigations in RE that are not yet representative. As a long-term goal, we aim at establishing an empirically sound basis for understanding practical trends and problems in RE, and for inferring representative improvement goals. Hence, the family of surveys will build a continuous and generalisable empirical basis for problem-driven RE research. 

Our family of surveys is designed and validated in joint collaboration with different researchers and practitioners from different countries. The instrument of the surveys further relies on a theory. This theory already integrates insights obtained from by now isolated empirical investigations in RE and contains a set of testable hypotheses. 

Furthermore, we reported on the results from the initial survey conducted in Germany where 58 respondents of different companies participated (with a response rate of 55\%). We tested the hypotheses in our theory and could corroborate various expectations ranging from ones about the status quo in RE as it is defined and lived in companies, the improvement of RE, and problems practitioners experience in their professional settings. 

Finally, we analysed our results for statistically significant correlations between the answers given to the different questions. The resulting correlations reveal, for example, that in cases practitioners stated to have flaws in their RE standard, they seem to have a multitude of flaws including an insufficient definition of roles and responsibilities, an unclear terminology, or an insufficient integration with further disciplines like risk management.

Our results already give us the opportunity to indicate to initial trends in RE as well as problems the respondents experience in their practical settings. Currently, we are in the planning phase of the first global replication conducted by various researchers in different countries, which constitutes the next step in bringing together the various empirical and RE-specific research communities. Each independent replication uses the same infrastructure and is based on the same questionnaire being modified in response to our results and our lessons learnt (see also the previous section). To guarantee a reproducible design of the survey and of the results, we commit ourselves to disclose the anonymised data of each replication to the PROMISE repository. The overall objective is to establish a regularly performed survey replication to continuously adapt and, finally, manifest a theory on the practical status quo and problems in RE (further information can be taken from \url{RE-Survey.org}).

In the following, we give a concluding summary of the results presented in this article and discuss those results in relation to existing evidence. We conclude with a brief discussion about the expected impact, the limitations of our study, and future work. 

\subsection{Summary of Results}

Our study revealed a broad set of findings presented in detail in Sect.~\ref{sec:Results} in relation to our theory. In the following, we give a brief summary of the results structured according to the research questions.

\paragraph{Expectations on good RE (RQ 1)}
Our survey showed that our respondents see RE as the discipline most beneficial, but also most challenging for an improvement stating that the missing organisational willingness to change would constitute the biggest barrier for such an improvement. When asked about the expectations on the RE standard, the results show a certain awareness for the importance of the role of RE artefacts. For example, our results indicate that participants see a clear definition of artefacts (and templates) as well as a clear terminology in their RE standard positive. They state the possibilities for a quality assurance of the created RE artefacts to be the biggest motivation to establish a company-wide RE standard. When defining such a standard, they rated agility to be the most important aspect the standard should support.

\paragraph{Definition, Application and Controlling of RE (RQ 2)}
Regarding the status quo in the RE standards, the respondents state to need more guidance on how to create precise RE artefacts. 45~\% of the respondents state their standard to include a description of expected RE artefacts and only 44~\% state that they have a clear description of roles and responsibilities. The application of the standard is in most cases (62~\%) performed by the project lead based on experiences while 53~\% rely on constructive quality assurance techniques. The status quo and the current problems the participants experience in their project environments might be also the reason why they expect their standard to support agility (see also the previous paragraph). 

\paragraph{Continuous Improvement of RE (RQ 3)}
Our results indicate that practitioners show a general reluctance against normative improvement approaches, such as CMMI. They mostly rely on a problem-driven improvement of their RE where company-specific goals and the organisational culture is rated as a reason against normative approaches. Exemplary statements for rejecting normative improvement approaches were ``We want to live our own agility'' or ``[because of the] individualism of the projects''. Whereas 55~\% of the respondents state an RE improvement to be managed via an own role or even an own business unit, 29~\% of the respondents state that an RE improvement is yet not done at all. 

\paragraph{Contemporary Problems and their Implications (RQ 4)}
We could reveal various problems in RE of which the highest ranked ones where incomplete requirements, moving targets, and time boxing -- all seen to be a main cause for experienced failed project. Further problems where communication flaws, missing traceability, terminological problems and unclear responsibilities. When asked what implications the problems have, we revealed that most of the stated problems seem to lead to change requests, additional communication and replanning, or to a stagnating process. We interpret, again, this result to speak for the need to better support agility in the RE.

\paragraph{Patterns in Expectations, Status Quo and Problems (RQ 5)}
We performed a correlation analysis among all variables in the complete data set and could reveal a set of statistically significant patterns. Those respondents who work in custom software development do, for example, not systematically improve their RE processes. We assume that the different environments and customer processes complicate a systematic improvement (see also the previous paragraphs). In the answers given to questions about the definition, application, and controlling of RE, we found that those respondents who think that offering standardised document templates and tool support benefits the communication also think that offering those templates increases the quality of the RE artefacts. We found a further set of correlations between the problems with the existing RE standard. An interesting result was that if a respondent has a problematic RE standard, it often offers many problems. For example, those who rate the RE standard as not sufficiently defining a clear terminology also rate it as giving no guidance on how to create the RE artefacts. Taking into account the results of the analysis of single questions where we found none of the problems to be significant, it thus seems either their RE standard has a multitude of problems or none.

\subsection{Relation to Existing Evidence}

Based on the statistical analyses of our survey results w.r.t.\ the theory presented in Sect.~\ref{sec:Theory}, the extension of this theory by applying Grounded Theory and a correlation analysis, we could confirm selected empirical findings available in literature. For example, although most respondents stated to have defined an RE standard, this standard is rated to have either no problems or a multitude of problems. The application of the standard and the tailoring of the model is stated to be done according to the expertise of the project lead. We interpreted those observation by tailoring being still an unsolved topic for RE~\cite{Mendez11}. Also, the results indicate that the definition of standardised artefacts in their RE standard is seen as important while leaving open the process to the participants, thus, supporting our own observations in the area of artefact orientation~\cite{MLPW11}.

The improvement of the RE standard is considered especially important (and challenging) for RE. Similar to the observations made by Staples et al.~\cite{SNJABR07} where smaller companies show a reluctance against normative improvement approaches, our results indicate to a more general reluctance independent of the company size. We interpret this as an agreement to the benefits seen in problem-driven RE improvement approaches~\cite{PIGO08, MW2013},  additionally supported by our observation that 79~\% of the respondents employ qualitative methods, for example, to gather lessons learnt.

Regarding the RE problems in practice, we could confirm the observations of, for example, Hsia et al.~\cite{HDK93} or Solemon, Sahibuddin, and And Ghani~\cite{Solemon:2009tf}. Those surveys reveal many problems that directly link to human factors such as communication problems, or inappropriate skills. We could extend their observation with a set of further problems, additionally ranked according to their frequency and relation for failed projects from the perspective of the respondents. For example, moving targets and time boxing were frequently stated problems experienced to have lead to failed projects. Our Grounded Theory could further extend those observations with consequences on the process including, inter alia, change requests or additional communication and replanning. 

Our results thus confirm some results from selected available studies and extend them with new ones.

\subsection{Impact/Implications}
We can directly infer two major implications from our contributions. First, our survey design relies on a theory we draw from different isolated studies and, thus, it integrates available studies to a broader view on the status quo in RE and its improvement as well as problems encountered in industrial environments. Our survey results further confirm parts of our theory and extends it with new concepts. This already gives us an initial saturation of the theory that allows researchers to steer their problem-driven research. They can define improvement goals on an empirical basis that already goes beyond isolated investigations and validation research; for instance, in the area of problem-driven RE improvement and in the area of artefact orientation. Second, the family of surveys is and will remain open. This allows not only researchers to reproduce the results and their interpretation, but also practitioners to evaluate their own RE situation against overall industrial trends.

\subsection{Limitations}
We have analysed the results from the survey conducted in Germany and the next replication round has already begun at the time of writing this article. Still, we are aware that our study still has limitations. Most importantly, our theory, although already integrating available empirical studies and extending their results with new ones, still needs to evolve and mature along with the variables in the instrument before we can finally demonstrate the ability to generalise from the results. Moreover, our results are based only on a reasonable but limited number of respondents from Germany.  We cannot make concrete statements about how generalisable the results are for Germany, let alone the rest of the world. On the contrary, we expect partially different results in different countries. For example, from our discussions with colleagues, we expect agile methods to be more influential in the Scandinavian countries. That means we need to follow our design of a family of surveys, conduct continuous replications in different countries and syntheses the results to establish a more reliable and empirically solid theory. 

Inherent to the nature of survey research is still that surveys can only reveal stakeholders' perceptions on current practices rather than empirically backed-up knowledge about those practices. To some extent, we aim at revealing exactly those perceptions (see, e.g., the research question on the expectations on good RE). For the questions about the current practices, we believe that we will mitigate this threat once we can demonstrate the ability to generalise as survey research allows us also to build a broad empirical basis. 

Furthermore, as we test a high number of hypotheses, the results from our hypotheses tests will suffer from the 
multiple comparisons problem. This means that because we run multiple tests, some will be tested as significant
by chance. There is an increase in the type I error but we chose not to correct for it to not increase the chance
for type II errors.

\subsection{Future Work}
We plan the further coordination of the replications, their syntheses, and their dissemination as future work. The dissemination aims at both empirical and RE research communities to support a variety of conceptual work in RE on basis of empirically sound findings. To this end, we cordially invite further researchers to join in for additional replications over the next years to establish an externally valid empirical basis on the state of the practice in requirements engineering. Furthermore, to have a more comprehensive basis for the underlying theory of the surveys, we plan to conduct a systematic literature review on expectations and the state of practice in RE.

\subsection{Acknowledgments}
We want to thank all participants of ISERN 2012 and EESSMod 2012 for fruitful discussions. In particular, we want to thank Silvia Abrah\~{a}o, Barry Boehm, Giovanni Cantone, Michael Chaudron, Maya Daneva, Marcela Genero, Lars Pareto, and Roel Wieringa for participating in the thematic workshops and giving us valuable feedback on the basic design of the family of surveys. Furthermore, we want to thank the internal and the external reviewers listed in Tab.~\ref{tab:review}. We are also thankful to Sebastian Eder and Saahil Ognawala for their support in the development of the coding scheme for the Grounded Theory during the initial run. Finally, we are grateful to Michael Mlynarski for supporting us in the pilot phase and to our project partners for their participation in the survey.

\bibliographystyle{model1-num-names}
\bibliography{napire}


\appendix

\section{Details to Axial Coding}
\label{sec:axialappendix}
\begin{figure}[!hbp]
\centering
  \includegraphics[width=0.9\textwidth]{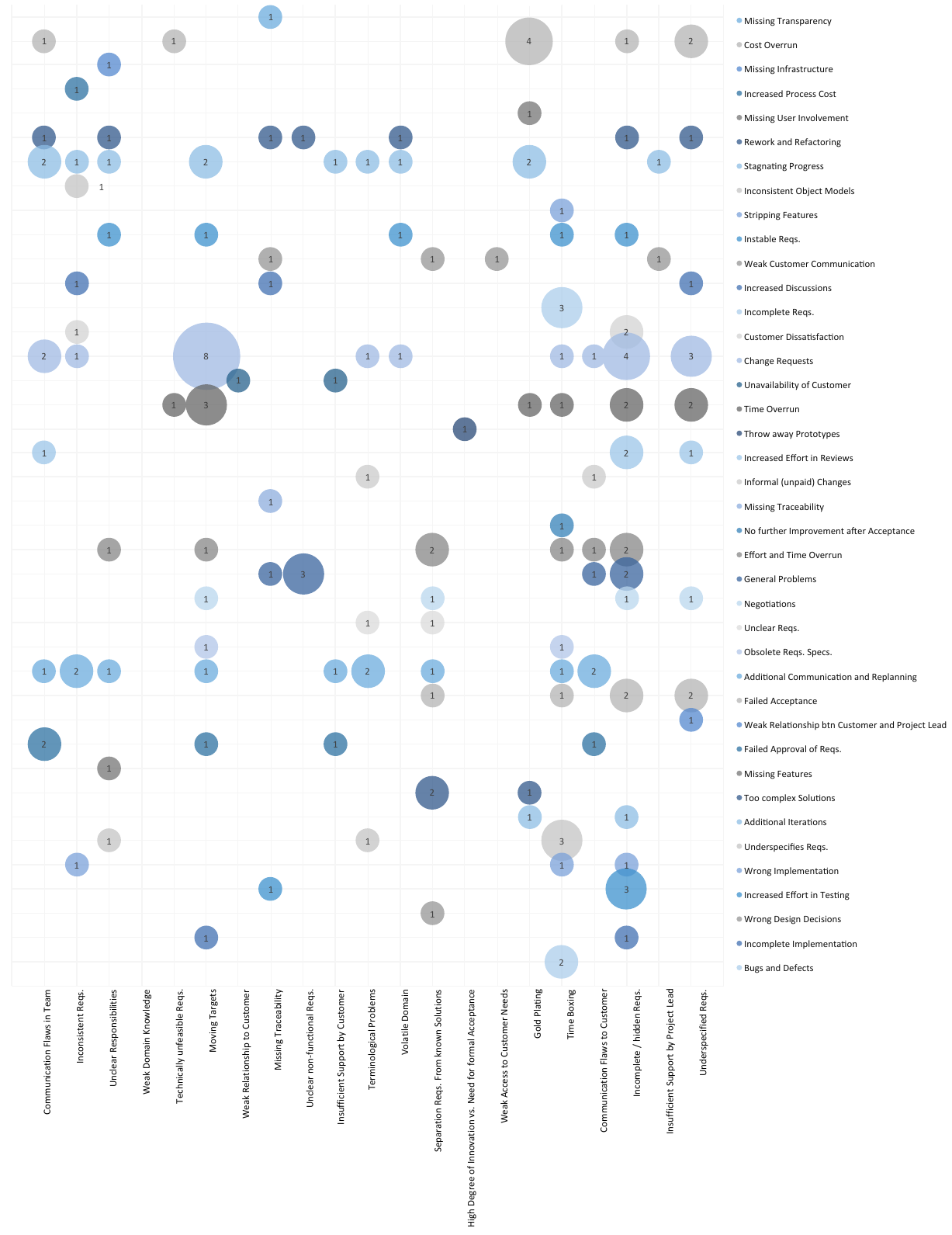}\\
  \caption{Axial coding: problems in relation to implications.}
  \label{fig.axialCoding1}
\end{figure}

The results of the axial coding for the consequences of given RE problems can be taken from Fig.~\ref{fig.axialCoding1}. In this figure, we illustrate for the given problem the number of dependencies stated by the respondents for the consequences. For instance, the problem ``Incomplete / Hidden Requirements'' is stated 4 times to result in change requests, or 3 times to result in an increased effort in testing.  Other examples include the problem ``Moving Targets'' resulting, inter alia, 8 times in change requests or the problem ``Inconsistent Requirements'' resulting in inconsistent object models.

The results of the axial coding for the reasoning for given RE problems is shown in Fig.~\ref{fig.axialCoding1}, which we organise the same way as the previous figure. Exemplary reasons stated, for example for incomplete / hidden requirements, were that no RE (phase) was explicitly planned as part of the tendering process. An exemplary suggestion for improvement  was to improve the artefact quality as one means to avoid incomplete requirements.

In contrast to the previous axial coding results, however, the results of the axial coding for the reasoning-related codes shows that -- except for two dependencies -- all dependencies were only stated one time. That means that the reasoning about the problems were only sporadically stated whereby we see those dependencies to suffer from an insufficient saturation during coding and the missing possibility for validation.

\begin{figure}[!hbp]
\centering
  \includegraphics[width=0.78\textwidth]{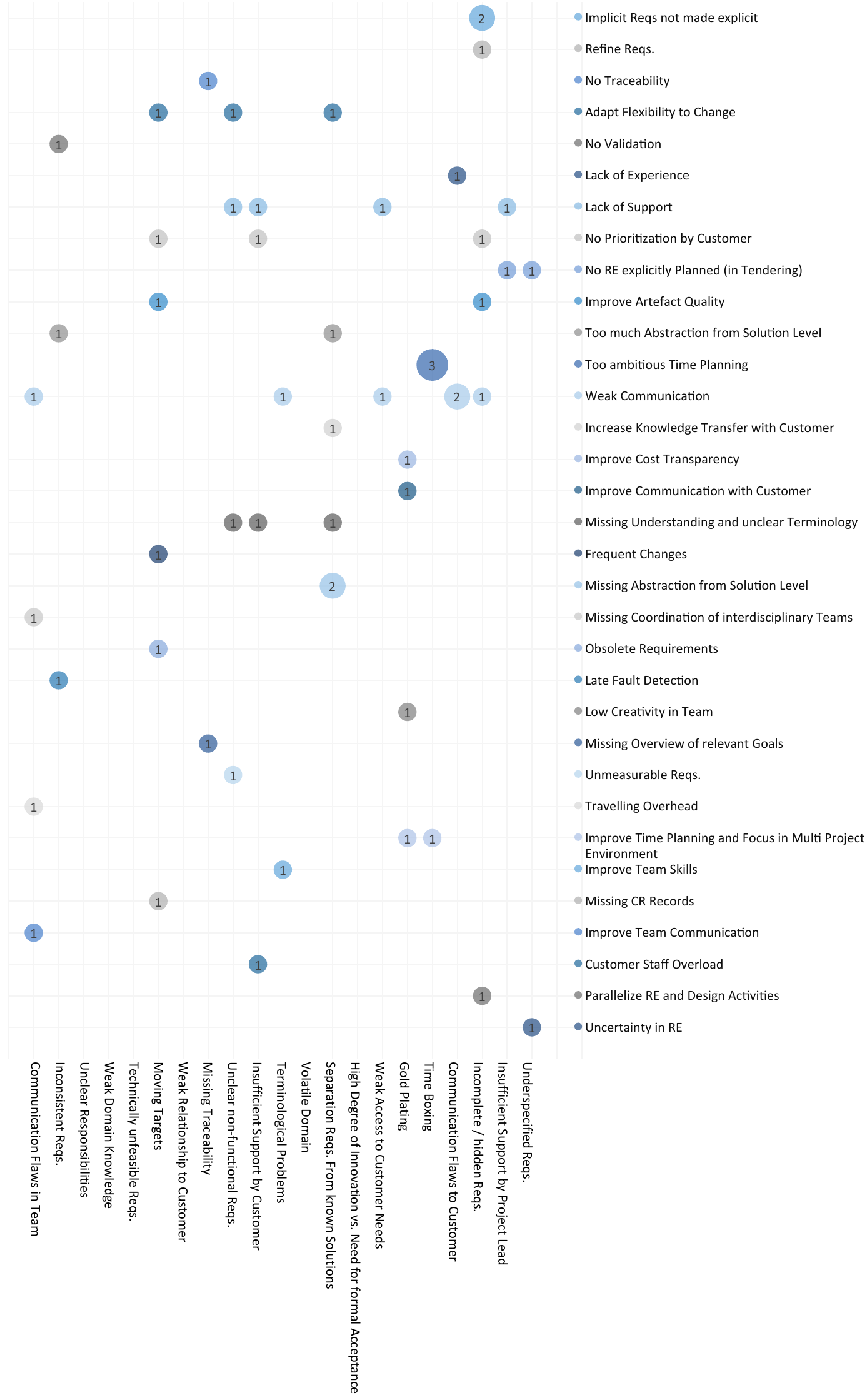}\\
  \caption{Axial coding: problems in relation to reasoning.}
  \label{fig.axialCoding2}
\end{figure}

\end{document}